\documentclass{sig-alternate}

\usepackage{color}
\usepackage[all,color]{xy}

\DeclareMathSymbol{\N}{\mathbin}{AMSb}{"4E}
\DeclareMathSymbol{\Z}{\mathbin}{AMSb}{"5A}
\DeclareMathSymbol{\R}{\mathord}{AMSb}{"52}

\newcommand{\reffig}[1]{Figure~\ref{#1}}
\newcommand{\refsec}[1]{Section~\ref{#1}}

\newcommand{\refeqn}[1]{Eqn. (\ref{#1})}

\newcommand{\ignore}[1]{}
\usepackage[titlenumbered,norelsize,ruled,vlined]{algorithm2e}
\usepackage{float}
\usepackage{subfigure}
\usepackage{hyperref}
\usepackage{boxedminipage}
\floatstyle{boxed}
\newfloat{Circuit}{H}{bpic}
\definecolor{RedLetter}{rgb}{0.63,0.165,0.163}
\definecolor{GreenLetter}{rgb}{0.165,0.63,0.163}

\begin{document}
\pagestyle{empty}
%
% --- Author Metadata here ---
\conferenceinfo{ICDT}{'14 Athens, Greece}
\setpagenumber{1}
\CopyrightYear{2014} % Allows default copyright year (2002) to be over-ridden - IF NEED BE.
%\crdata{0-12345-67-8/90/01}  % Allows default copyright data (X-XXXXX-XX-X/XX/XX) to be over-ridden.
% --- End of Author Metadata ---

\title{Transaction Repair: Full Serializability Without Locks}
%
% You need the command \numberofauthors to handle the "boxing"
% and alignment of the authors under the title, and to add
% a section for authors number 4 through n.

\numberofauthors{1}
\author{
% 1st. author
\alignauthor
Todd L. Veldhuizen\\
   \affaddr{LogicBlox Inc.}\\
   \affaddr{Two Midtown Plaza}\\
   \affaddr{1349 West Peachtree Street NW}\\
   \affaddr{Suite 1880, Atlanta GA 30309}\\
%   \email{\small {\sf tveldhui@\{logicblox.com,acm.org\}}}
}
\maketitle
\begin{abstract}
Transaction Repair is a method for lock-free, scalable transaction
processing that achieves full serializability.  It demonstrates
parallel speedup even in inimical scenarios where all pairs of
transactions have significant read-write conflicts.  In the transaction
repair approach, each transaction runs in complete isolation in a
branch of the database; when conflicts occur, we detect and repair
them.  These repairs are performed efficiently in parallel, and the
net effect is that of serial processing.  Within transactions, we
use no locks.  This frees users from the complications and performance
hazards of locks, and from the anomalies of sub-serializable
isolation levels.  Our approach builds on an incrementalized variant
of leapfrog triejoin, an algorithm for existential queries that is
worst-case optimal for full conjunctive queries,
and on well-established techniques from programming
languages: declarative languages, purely functional data structures,
incremental computation, and fixpoint equations.

\end{abstract}

%A category including the fourth, optional field follows...
%\category{D.2.8}{Software Engineering}{Metrics}[complexity measures, performance measures]

%\terms{Algorithms,Theory}

%\keywords{ACM proceedings, \LaTeX, text tagging} % NOT required for Proceedings

\section{introduction}

\subsection{Scenario}

\label{s:scenario}

Consider the following artificial scenario chosen to
highlight essential issues.
A database tracks available quantities of warehouse items identified by sku number 
(\emph{stock-keeping unit}).
Each transaction adjusts quantities for a subset of skus,
updating a database predicate $\mathsf{inventory}[\mathsf{sku}]=\mathsf{qty}$.
Suppose there are $n$ skus, and each transaction adjusts 
skus chosen independently with probability
$\alpha n^{-1/2}$.
Most pairs of 
transactions will conflict when $\alpha \gg 1$: the expected number of skus common to
two transactions is $E[\cdot] = n \cdot \left( \alpha n^{-1/2} \right)^2 = \alpha^2$,
an instance of the Birthday Paradox.

\begin{figure*}
\includegraphics[width=\columnwidth]{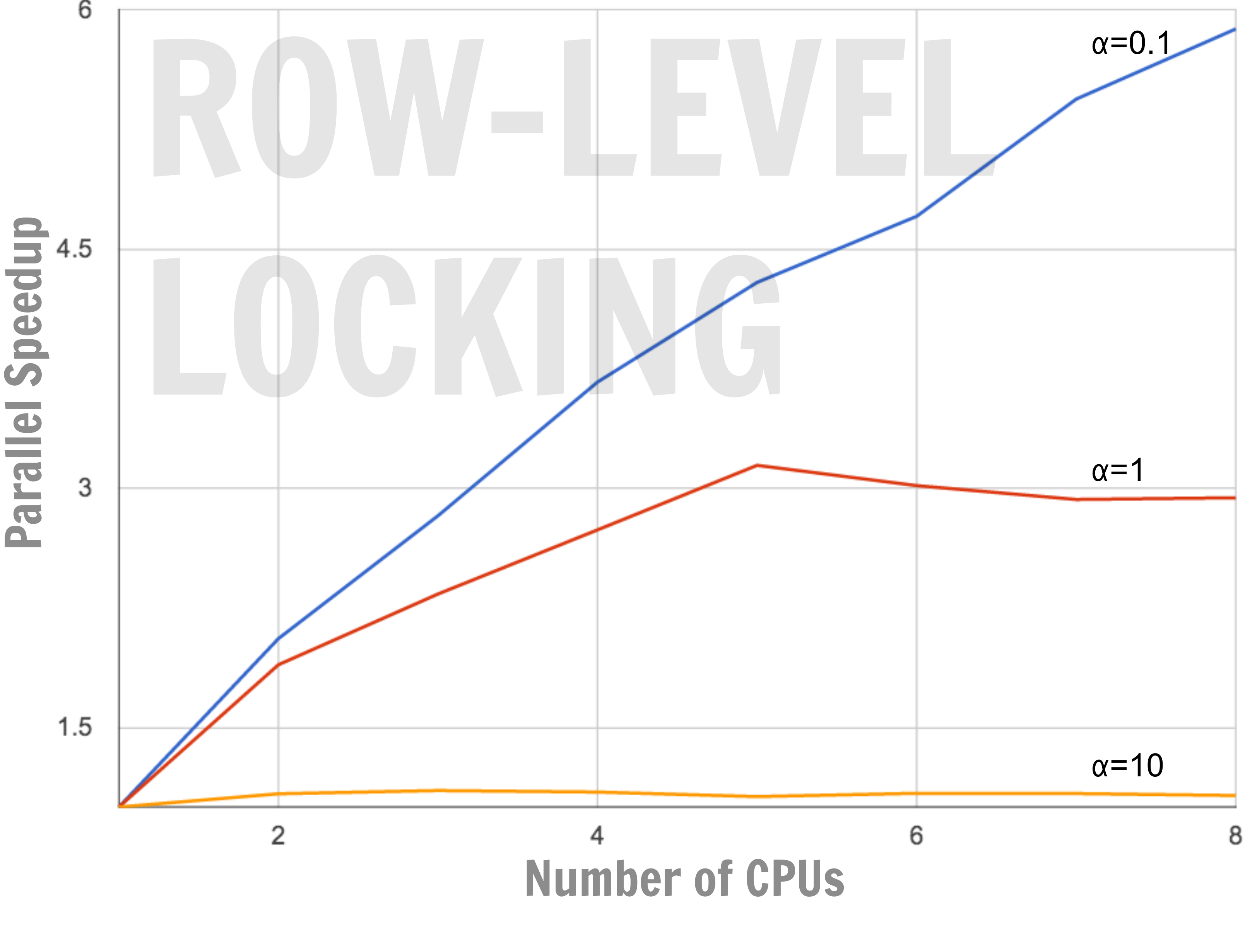}
\includegraphics[width=\columnwidth]{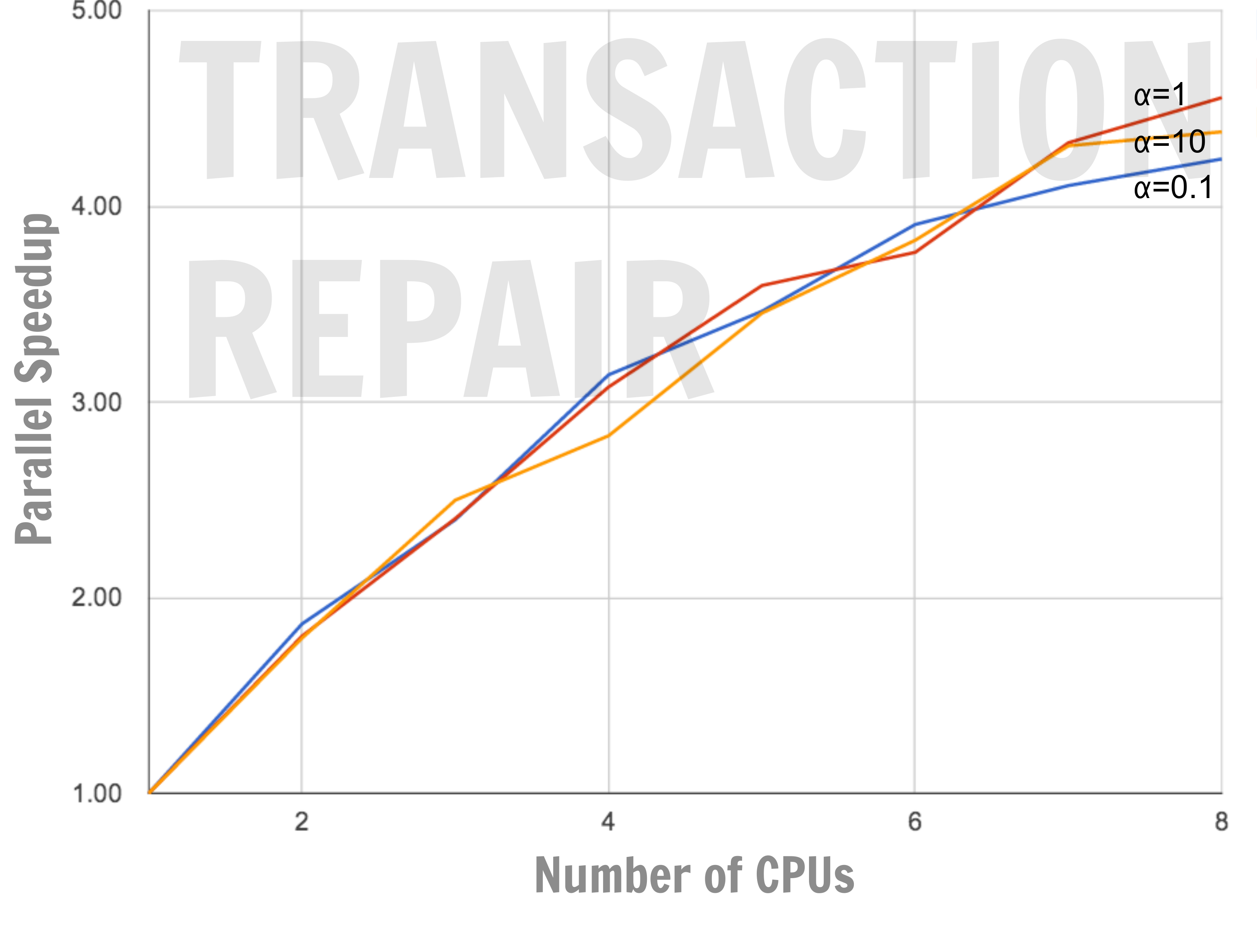}
\caption{
\label{f:speedup}
(a) Parallel speedup for the scenario of \refsec{s:scenario}, using
an efficient implementation of row-level locking.  When the expected number
of conflicts between two transactions is small ($\alpha=0.1$), parallel
speedup for a modest number of processors is possible, since most transactions
can get some work done before encountering a lock conflict.  For $\alpha=1$, parallel
speedup is sharply limited; and for $\alpha=10$ almost no speedup is possible,
since every transaction immediately encounters a lock conflict.
(b) Parallel speedup for the same scenario, using transaction repair.
(Speedup of our current prototype is hindered by contention in
memory allocation subroutines; this is unrelated to repair per se.)
}
\end{figure*}

Row-level locking \nocite{Bernstein:CSUR:1981} is a bottleneck when $\alpha \gg 1$: since most transactions
have skus in common, they quickly encounter lock conflicts
and are put to sleep.  \reffig{f:speedup} (left) shows parallel speedup of
transaction throughput for $\alpha=0.1$, $\alpha=1.0$, and $\alpha=10$,
using an efficient implementation of row-level locking on a multicore machine.
Note that for $\alpha=10$ there is no parallel speedup: there are so many
conflicts that throughput is reduced to that of a single cpu.

Our approach, which we call \emph{transaction repair}, is rather different.
The LogicBlox database has been engineered from the ground-up to use
purely functional and versioned data structures.
Transactions run simultaneously, with no locking, each in complete isolation
in its own branch of the database.  We then detect conflicts and repair them.  These repairs
are performed efficiently in parallel, and the net result is a database
state indistinguishable from sequential processing of transactions.
With this approach, we are able to achieve parallel speedup even when
there are large amounts of conflicts between
transactions (\reffig{f:speedup}, right).

It does not strain credulity to report that transaction repair can achieve
parallel speedup for the trivial scenario just described.  Remarkably, our technique
applies to arbitrary mixtures of complex transactions.

\subsection{Transaction repair}

Transaction repair combines three major ingredients:
\begin{enumerate}
\item Leapfrog triejoin: Each transaction in our system consists of one
or more \emph{rules} written in our declarative language LogiQL, a substantial
augmentation of Datalog which preserves the clean lines of the original.
Each LogiQL rule is evaluated using
\emph{leapfrog triejoin}, an algorithm for
existential rules for which a significant optimality property was
recently proven \cite{Veldhuizen:ICDT:2014}.
\item Incremental maintenance of rules: Leapfrog triejoin admits an efficient
incremental maintenance algorithm
that is designed to achieve cost proportional to the trace-edit
distance of leapfrog triejoin traces \cite{Veldhuizen:LB:2013}.
We employ this algorithm to repair individual
rules when conflicts occur between transactions.  In operation,
the maintenance algorithm collects \emph{sensitivity indices} that
precisely specify database state to which a rule is sensitive, in the sense
that modifying that state could alter the observable outcomes of the transaction.
Maintenance of individual rules is extended to maintenance of entire transactions
by propagating changes through a dependency graph of the transaction rules
(\refsec{s:graph}).
\end{enumerate}

The third ingredient is transaction repair circuits, which we broadly outline in
\refsec{s:repair}, and describe in detail in subsequent sections.

A bottom-up exposition would begin at the level of single rules and
leapfrog triejoin, and describe how transaction repair is built on these
foundations.  However, the novelty of this paper is in
the higher-level aspects of transaction repair;
we begin there since the principles can be understood
while abstracting away the detailed mechanics of individual transactions.
We return to leapfrog triejoin and its incremental maintenance
in \refsec{s:lftj}.  In \refsec{s:graph} we cover the middle ground
between rules and transaction repair, namely, the repair mechanisms
that are employed inside the boundaries of a single transaction.

\subsection{Advantages}

Transaction repair promises significant advantages over existing approaches
to concurrency control:

{\bf Simplicity.} We present the simplest possible concurrency
model to users: from their vantage point, transactions behave as if processed
one at a time, but they enjoy the performance benefits of parallelism.
There are no locks, and hence no lock interactions, no transactions
aborted due to lock conflicts, and no locking strategies to select or tune.
Since transaction repair provides full serializability, users are freed from
the anomalies, hazards, performance tradeoffs and anxieties of
sub-serializable isolation levels.
Unlike Multi-Version Concurrency Control (MVCC) \cite{Bernstein:CSUR:1981},
we only abort transactions if integrity constraints fail.

{\bf Performance.}  We are able to achieve parallel speedup even in
inimical scenarios, for example, all pairs of transactions having
significant conflicts.  This improves on Optimistic Concurrency
Control (OCC) \cite{Kung:TODS:1981}.  OCC employs \emph{readsets},
somewhat analogous to our \emph{sensitivity signals}.  However,
when conflicts are detected, OCC restarts transactions, rather than
repairing them as we do.  This causes OCC to perform poorly when there
is significant conflict between transactions.

{\bf Economics.}  Our approach to transaction processing is inherently scalable.
Since we do not use locks for concurrency control, we do not require low-latency
communication between compute nodes to confer over locks and versioning.
This suggests that transaction repair may be able to achieve transaction
throughput on inexpensive clusters that rivals that of traditional databases
on high end hardware.

\subsection{Transaction Repair Circuits}

\label{s:repair}

We introduce the basic concepts of transaction
repair at a casual level of detail, with pointers to later sections
where the concepts are developed in depth.

\begin{figure}[t]
\centering
\subfigure[Single transaction as a black box.]{\label{f:twotxns-a}
\includegraphics[scale=1.0]{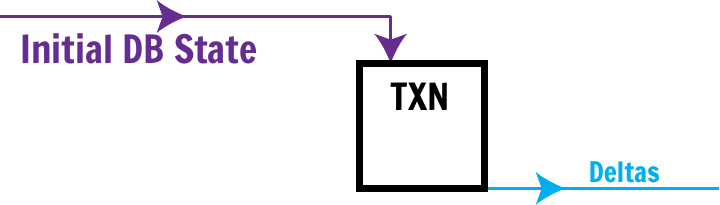}
}\\
\subfigure[Two transactions run in parallel, with corrections from the first used to repair the second.]{\label{f:twotxns-b}
\includegraphics[width=\columnwidth]{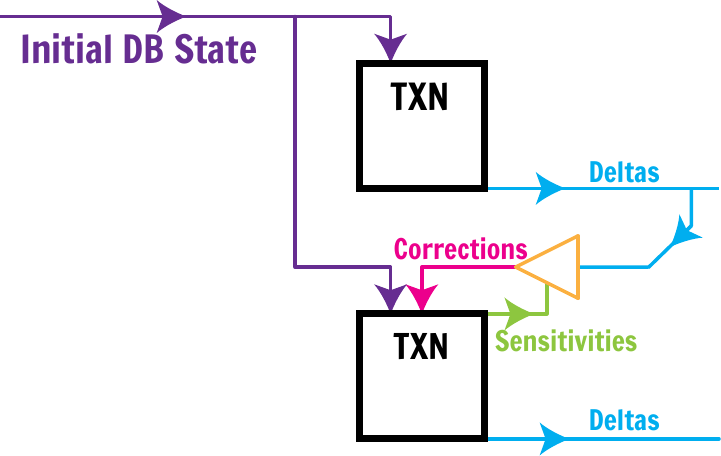}
}
\caption{\label{f:twotxns}Some basic repair circuits.}
\end{figure}

We can view a transaction as a black box which inputs an initial
database state and outputs changes to the database we call
\emph{deltas} (\reffig{f:twotxns-a}).  Suppose we run two transactions
independently in parallel.  We have to be concerned that the second transaction
in the proposed serialized order tries to read some state affected
by deltas of the first transaction.   To address this, transactions
report their \emph{sensitivities}, that is, aspects of database state
whose modification might alter the outcome of the transaction.
We compare the deltas produced by the first transaction
to sensitivities declared by the second, to test whether there is
a possibility of conflict.  If so we describe relevant \emph{corrections} of
the database state (\reffig{f:twotxns-b}) to the second transaction,
which is then repaired (incrementally maintained) for the
corrections (Sections \ref{s:lftj}, \ref{s:graph}).

\reffig{f:twotxns-b} is a simple example of a \emph{transaction repair circuit}.
It is not a circuit in the sense of custom hardware, but rather a schematic describing the
work to be performed; in particular, it specifies a set of recursive
fixpoint equations to be solved (\refsec{s:running}).
The deltas, corrections, and sensitivities are
\emph{signals} (\refsec{s:signals}).  
The triangle element is a \emph{correction operator}: it takes
as inputs changes to the database state (in this case, deltas) and declared
sensitivities, and selects just those deltas that
match sensitivities (\refsec{s:corr}).

\begin{figure}[t]
\centering
\subfigure[]{\label{f:twotxns-c}
\includegraphics[width=\columnwidth]{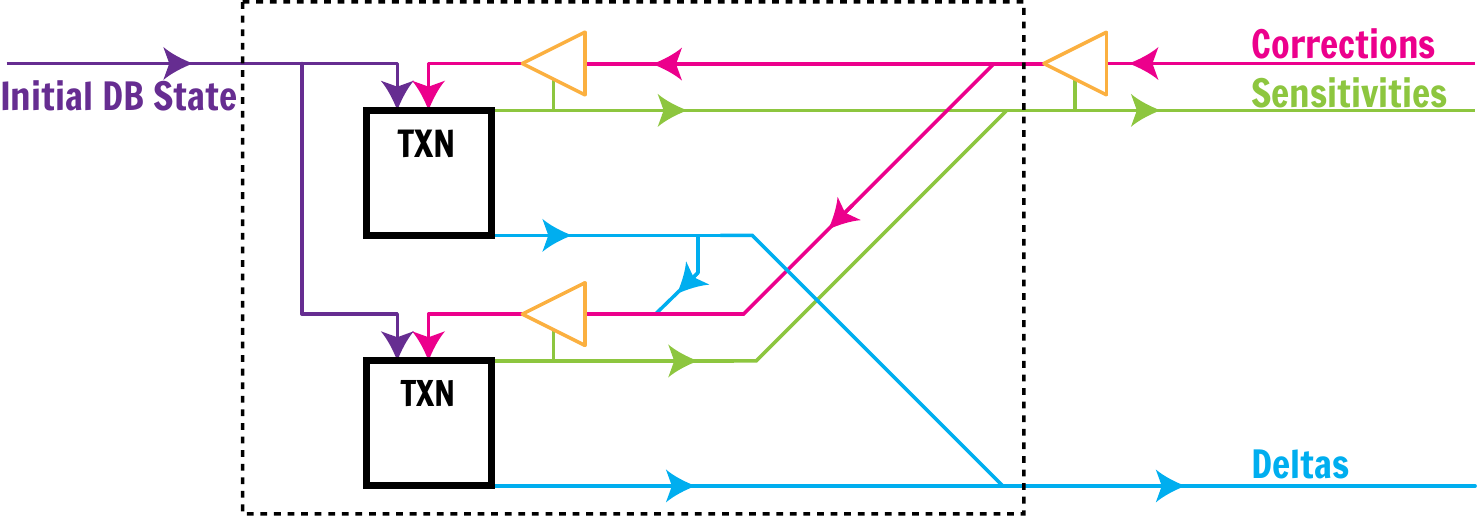}
}\\
\subfigure[]{\label{f:twotxns-d}
\includegraphics[width=\columnwidth]{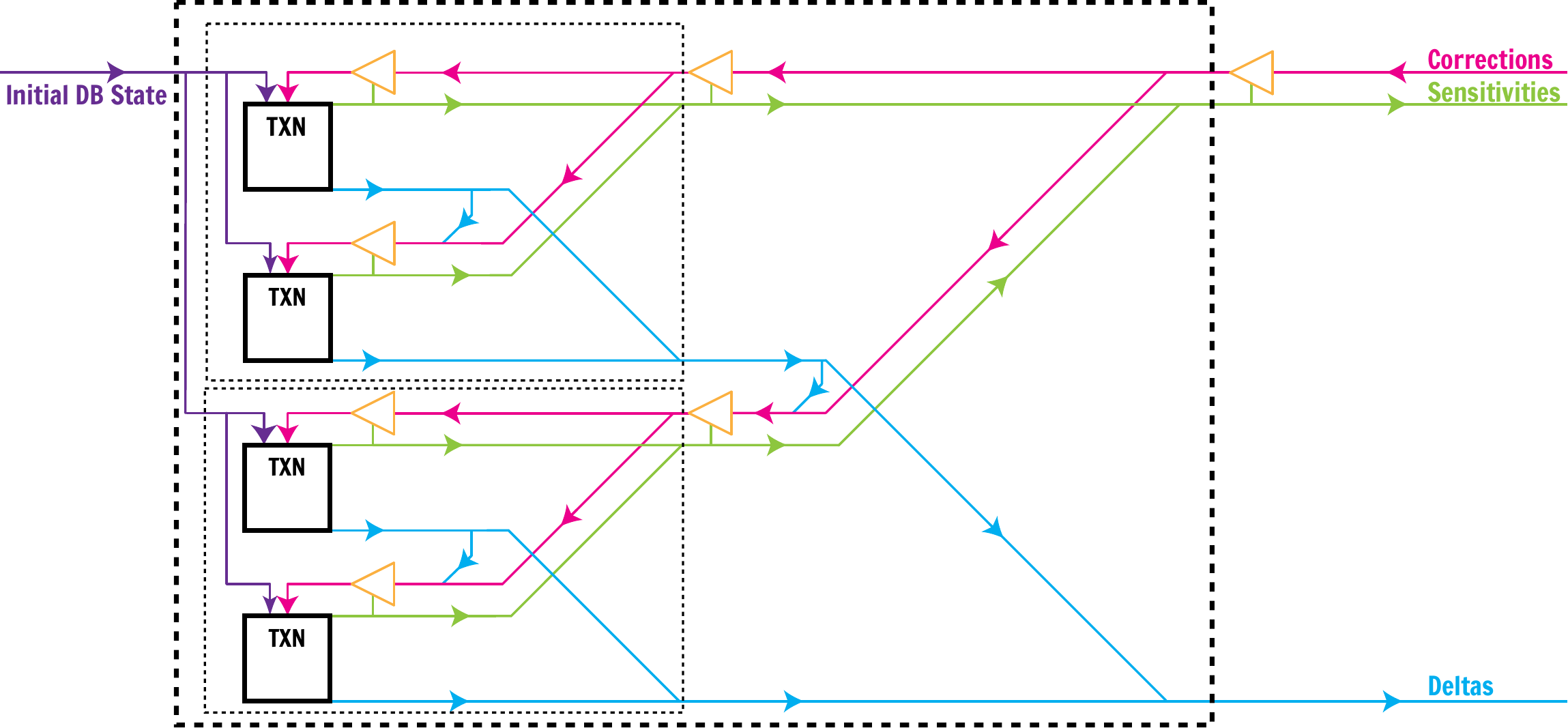}
}
\caption{\label{f:twotxnscd}Repair circuits for (a) two and (b) four transactions.}
\end{figure}

Suppose there are previous transactions before these two in the serialization order.
We need to know what
corrections might apply to these two transactions.  In
\reffig{f:twotxns-c}, we have the first
transaction also report its sensitivities; we merge the sensitivities
of the two transactions, and add correction circuitry that filters
incoming corrections and feeds them back to repair the transactions.
To determine the net changes made by the two transactions,
we merge their deltas, giving priority to changes
made by the second transaction: since it occurs later in the serialized
order, its changes supersede those of the first.  The merging of the  
deltas is accomplished by a $\mathsf{merge}$ operator (\refsec{s:merge}).

Consider the dotted box of \reffig{f:twotxns-c}.  Outside
this box we observe the same structure as a single transaction:
initial database state and corrections enter the box, and
sensitivities and deltas come out.  In \reffig{f:twotxns-d} we duplicate that
box of two transactions and wire the boxes as done in \reffig{f:twotxns-c} for
individual transaction boxes.
This constructs a transaction repair circuit for four transactions; iterating
yields repair circuits for $(2^k)_{k \in \omega}$ transactions.
Each of the dotted boxes of \reffig{f:twotxns-d} represent a
transaction \emph{group}; individual transactions and groups are
arranged in a \emph{transaction tree} (\refsec{s:txntree}).

\begin{figure}[t]
\includegraphics[width=\columnwidth]{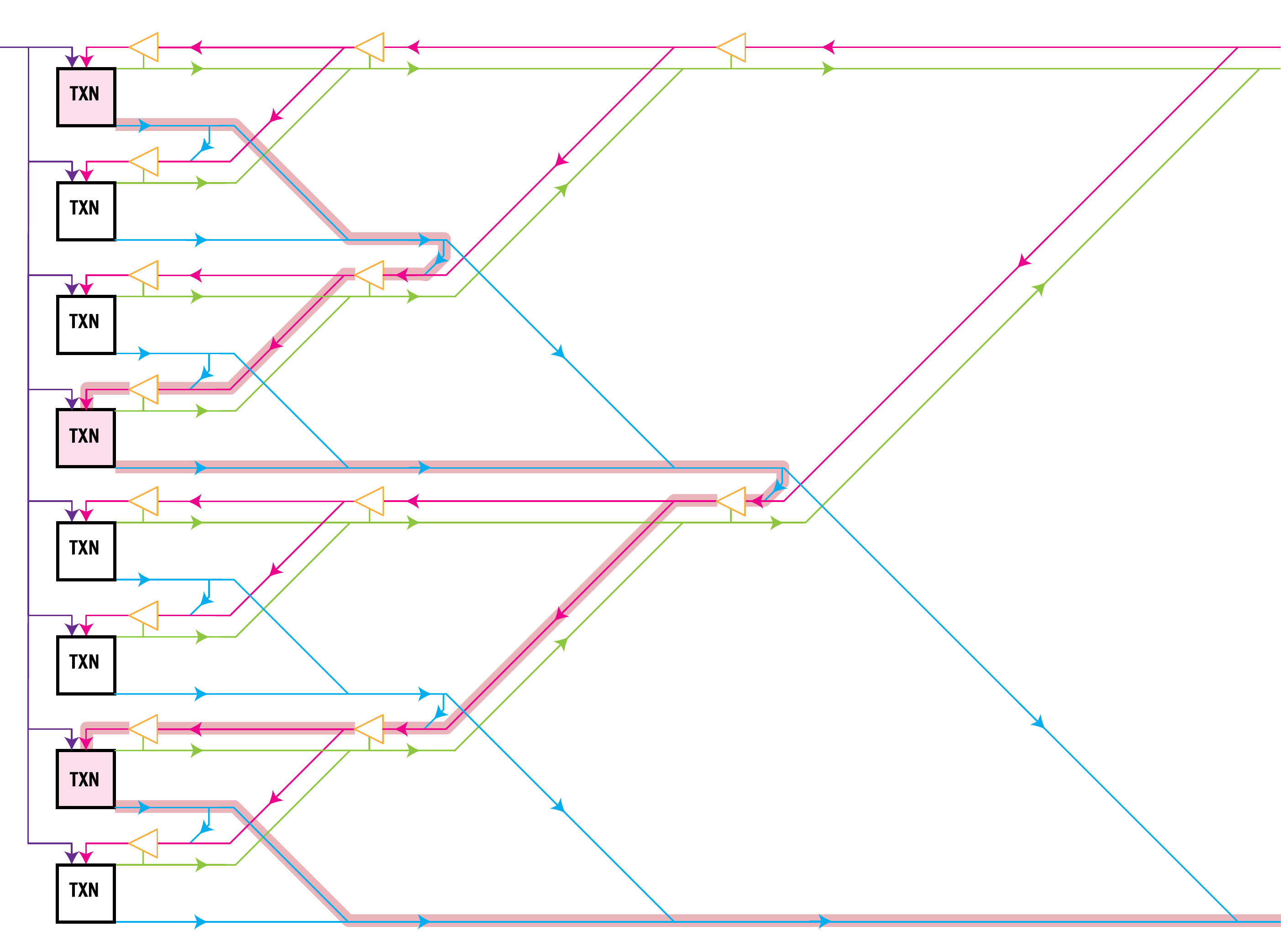}
\caption{\label{f:conflictchain}A conflict chain of three transactions.
Deltas from the first transaction are propagated as corrections to the
fourth, and deltas from the fourth are propagated as corrections to the
seventh transaction.}
\end{figure}

Conceptually, repair circuits are evaluated by fixpoint iteration
(\refsec{s:running}):
when inputs to a merge operator, correction operator, or
individual transaction change,
the circuit element is \emph{refreshed}, and its outputs
are revised as appropriate.
For simple transactions, the number of steps
required for convergence is controlled by the length of the largest
\emph{conflict chain}.
% Interesting question: for what complexity classes is this true?
Multiple transactions both reading and writing the same record
are a common cause of conflict chains.
\reffig{f:conflictchain} illustrates a conflict chain among
three transactions: the first transaction modifies a record
read then written by the fourth transaction; the seventh transaction
reads the record.
As the fixpoint iteration proceeds, deltas from the first
transaction are routed as a correction to the next transaction
in the conflict chain, and its deltas will be routed to the next
transaction in the conflict chain, until all the conflicts are
resolved and the combined deltas are consistent with serial
execution.

\begin{figure}[t]
\includegraphics[width=\columnwidth]{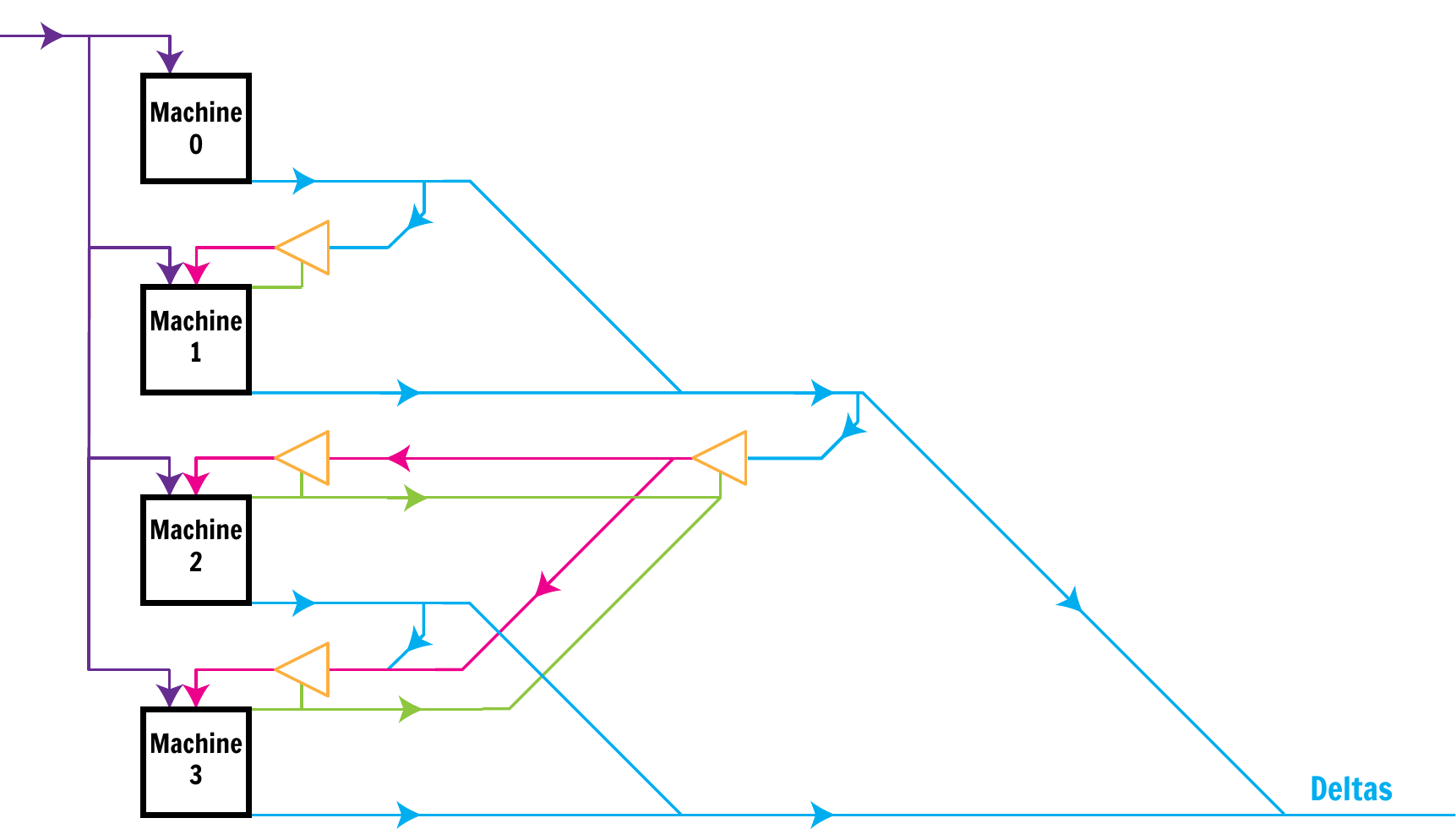}
\caption{\label{f:FourMachines}Communication circuit for
transaction repair on a four-machine cluster.}
\end{figure}

The repair circuit diagrams expose parallelism in an obvious way:
multiple cpus can be simultaneously evaluating and repairing transactions,
and computing merges and corrections.
We need not limit ourselves to multiple cores.
If transaction arrivals exceed the capacity of a single machine,
we can scale out to clusters:
label the outermost box of \reffig{f:twotxns-d} `machine 0';
repeating the earlier constructions
we obtain a repair circuit for four machines (\reffig{f:FourMachines}).
The signal lines carrying deltas, sensitivities, and corrections
become communications between machines.
In this diagram we have omitted correction and sensitivity signals
that would be unnecessary, assuming machine 0 contains the first
transactions in a commit group.

% Kung, H.T. (1981). "On Optimistic Methods for Concurrency Control". ACM Transactions on Database Systems
\nocite{Kung:TODS:1981}

% Principles of Transaction Processing, Bernstein
\nocite{Bernstein:2009}

% Materialized Views: Techniques, Implementations, and Applications
\nocite{Gupta:1999}

% Optimistic Concurrency Control by Melding Trees
% http://www.vldb.org/pvldb/vol4/p944-bernstein.pdf
\nocite{Bernstein:VLDB:2011}

\section{Repair circuits in depth}
\label{s:circuits}

We now describe repair circuits in depth.

For simplicity, the exposition of \refsec{s:repair} omitted a
major detail.  From \reffig{f:twotxns-d} one might construe that
the largest dotted box emits a single signal containing changes
from four transactions, and in general the work performed by
top-level merge operators is proportional
to the number of transactions.  This is not the case.

\begin{figure}[t]
\includegraphics[width=\columnwidth]{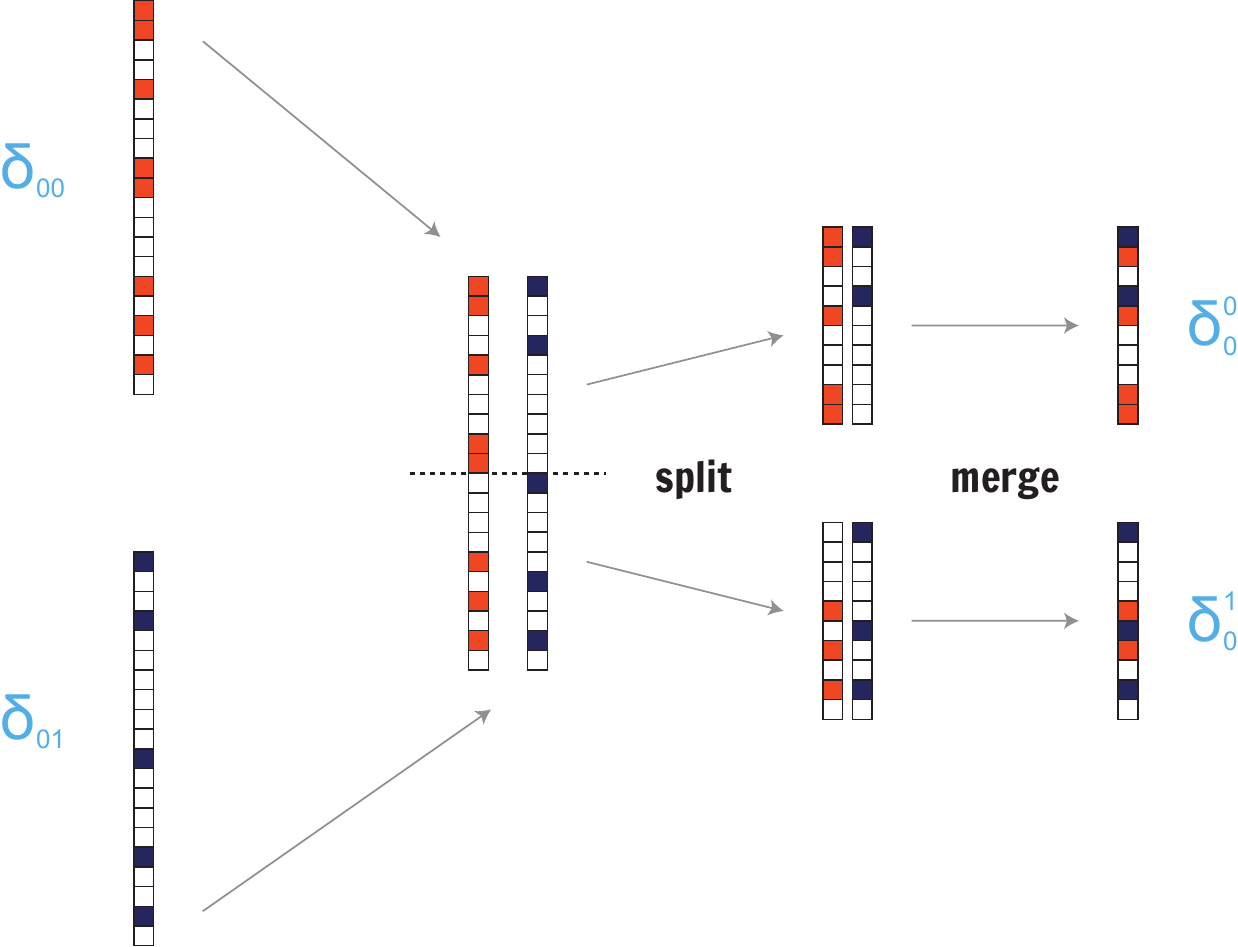}
\caption{\label{f:splitmerge}Operation of a merge operator.  Delta signals $\delta_{00}, \delta_{01}$
from two transactions are partitioned into subdomains at a split point.  Delta records below the
split point are placed in $\delta_0^0$, and the others in $\delta_0^1$.}
\end{figure}

To better expose parallelism, when we merge delta signals,
we simultaneously perform a \emph{domain split}, separating the merged
deltas into those for two disjoint subdomains.
\reffig{f:splitmerge} illustrates the merge of two
delta signals $\delta_{00}, \delta_{01}$.
The subscripts are a binary representation of the transaction
label (zero and one in a group of four transactions),
indicating the order of transactions in the serialization order.
Records describing changes
to the first half of the domain are placed in a signal $\delta_0^0$,
and the remainder are placed in $\delta_0^1$.  The subscript $0$
of these signals labels the transaction \emph{group}; a group
with label $t$ contains all transactions whose label begins with
the prefix $t$.  The superscript labels the domain split: the
two halves of the domain are given subdomain labels $0$ and $1$, and when
split into quarters the subdomains would be labelled $00$, $01$, $10$,
and $11$.  We will refer to subdomain labels using the letter $d$,
and transaction group labels using the letter $t$.  A merge operator
for deltas always takes a pair of signals of the form
$\delta_{t0}^d$ and $\delta_{t1}^d$, which carry deltas in a subdomain
$d$ for two sibling transaction groups $t0$ and $t1$, and emits
a pair of signals with form $\delta_t^{d0}$ and $\delta_t^{d1}$:
these are deltas for the two halves of the subdomain $d$
(namely, $d0$ and $d1$) for the entire group of transactions
$t$ (which contains all transactions in groups $t0$ and $t1$).

\reffig{f:domainsplit} shows the detailed wiring of delta and
sensitivity signals for a group of four transactions.  The
four delta signals emitted by individual transactions are
merged into four delta signals, each corresponding to a
quarter of the domain.   Merging of sensitivity signals is
similar to that of delta signals, with modest adjustments
(\refsec{s:merge}.)

%%%%%%%%%%%%%%%%%%%%%%%%%%%%%%%%%%%%%%%%%%%%%%%%%%%%%%%%%%%%%%%%%%%%%%%%%
\subsection{Domains and subdomains}

We now define \emph{the domain}.
In brief, all tuples in the database
can (conceptually) be placed in one set $D$ with a natural total order
$(D,\leq)$.   We specify subdomains of the database as
intervals $[d_1,d_2)$ where $d_1,d_2 \in D$.  Tuples of $D$
have form $(P,\overline{k})$, where $P$ is
a relation or function symbol, and $\overline{k}$ is a key tuple.
In practice fixed-width identifiers (e.g. integers) can be used to
label predicates.  Readers content with this synopsis of $D$
may skip to the next heading,
bypassing some mildly belabored definitions.

The LogicBlox database engine implements a key-value store with
set semantics.  A database instance has much in common with
a structure of (many-sorted) first-order logic.
A database schema defines
function symbols $F_0$, $F_1$, $\ldots$, $F_m$ and relation symbols
$R_0$, $R_1$, $\ldots$, $R_n$, each with an associated signature
specifying key and value arities, and types.
(For example,
$F[\mathsf{int}$, $\mathsf{string}]=\mathsf{decimal}$
specifies a function with key-arity 2 and value-arity 1.)
A nullary function (with key arity 0) is a scalar;
relations always have value arity 0.  Each type $T$ has
an associated total order $(T,\leq_T)$ predefined and/or
dynamically maintained by the database engine.

We refer to functions and relations as
\emph{predicates}.  For most purposes of this paper,
a predicate symbol $P$ has an associated 
data structure representing the elements of $P$.
(Predicate symbols can also represent \emph{primitives}
such as integer addition, multiplication, etc.; these
are largely irrelevant to our purpose, apart from a brief mention in
\refsec{s:lftj}.)

For each predicate P, let $T_P = T_1 \times \ldots T_k$,
where $T_1,\ldots,T_k$ are the key types of its signature,
and let $(T_P,\leq_{T_P})$ be the lexicographical order
on tuples.
The data structure for a predicate $P$
provides $O(\log |P|)$ lookups, and $O(1)$ amortized
iteration of elements in the order $\leq_{T_P}$.

Finally, we define the domain $D$ by:
\begin{align}
D &\equiv \{ +\infty, -\infty \} \cup \left( \bigcup_P \{ P \} \times T_P \right)
\end{align}
where $\pm \infty$ are endpoints of the domain,
and $\bigcup_P \cdot$ ranges over function and relation symbols.
Elements of $D$ other than $\pm\infty$
are of the form $(P,\overline{k})$,
where $P$ is a predicate symbol, and $\overline{k}$ is a key tuple.
Fix an arbitrary order on predicate symbols,
and define a total order $(D,\leq_D)$ where
$\pm\infty$ are the endpoints,
and lexicographical order on elements of the form
$(P,\overline{k})$: tuples
ordered first by predicate symbol, and secondarily
by key order within each predicate.
Subdomains can then be specified as intervals
$[d_1,d_2)$, where $d_1,d_2 \in D$.

%%%%%%%%%%%%%%%%%%%%%%%%%%%%%%%%%%%%%%%%%%%%%%%%%%%%%%%%%%%%%%%%%%%%%%%%%
\subsubsection{Domain decomposition}

\label{s:domaindecomp}

A domain decomposition is specified by a binary
tree, each node labelled by a tuple
$s \in D$ of the form $(P,\overline{k})$,
$P$ being a predicate symbol.
The tuple $s$ is a \emph{split point}.
The root node
of the tree is associated with the subdomain $[-\infty,+\infty)$
of $D$.  A root node with split point $s$ has a 
left child ssociated with the subdomain $[-\infty,s)$ and a right child
with $[s,+\infty)$; and so on down the tree.
We specify a subdomain with a binary string $d \in \{0,1\}^\ast$
identifying a tree path,
starting from the root and taking left or right
branches for 0 or 1, respectively.
Split points can be placed between predicate symbols
by choosing something with form $(P,[-\infty,\ldots,-\infty])$,
or within the records of a predicate.

For exposition it is handy to assume the
domain decomposition tree has height $\log_2 n$, where
$n$ is the number of transactions in the repair circuit
under discussion.  In practice the height and contents
of the domain decomposition tree can be tweaked to finesse performance.
For example, if there is an influx of microtransactions,
one might elect to defer domain splitting until a group
of transactions is large enough to warrant it.  This
can be accomplished by dynamically revising
a domain decomposition subtree specifying subdomains
$[s_0,s_1), [s_1,s_2), \ldots, [s_{k-1},s_k)$ into
one with subdomains $[s_0,s_k), \emptyset, \ldots, \emptyset$.

%%%%%%%%%%%%%%%%%%%%%%%%%%%%%%%%%%%%%%%%%%%%%%%%%%%%%%%%%%%%%%%%%%%%%%%%%

\subsection{The transaction tree}

\label{s:txntree}

Individual transactions are arranged in a serialization order, and
labelled with binary strings indicating their placement: $000$, $001$, $010$, $\ldots$
$111$ for a group of eight.
We'll say a transaction is \emph{later} than another if its label is
lexicographically greater.  We will use the symbol $t$ to represent
a transaction label or a prefix of a transaction label.

\begin{figure}[t]
\includegraphics[width=\columnwidth]{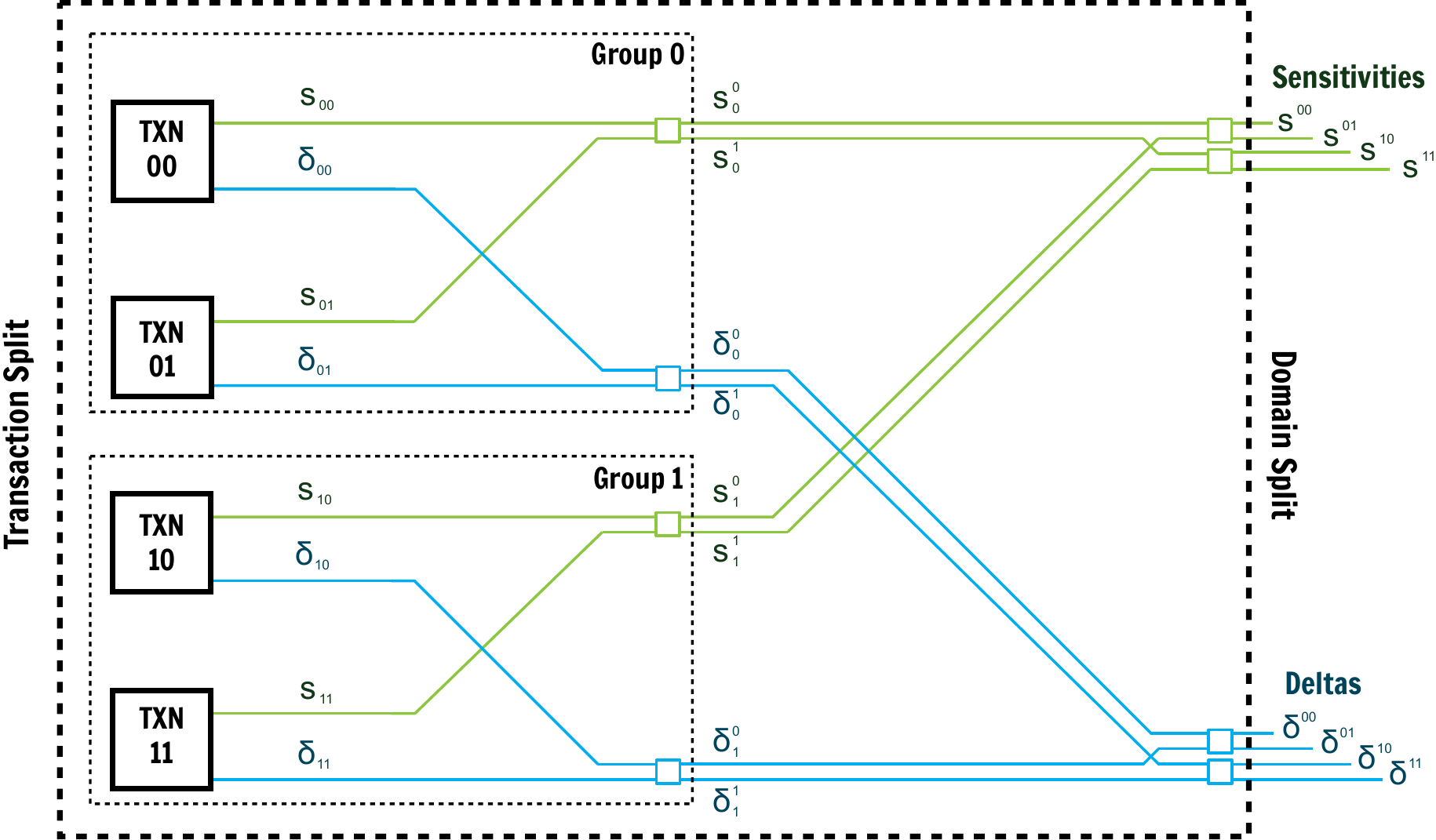}
\caption{\label{f:domainsplit}Transaction split and domain split.  Small blue boxes indicate
$\mathsf{merge}$ operators, and green boxes are $\mathsf{corr}$ (correction)
operators.}
\end{figure}

The \emph{transaction tree} is a binary tree whose leafs are transactions,
and internal nodes are transaction \emph{groups}.
Groups nodes are labelled in the obvious way: a group node
with label $t$ has \emph{left child} $t0$ and possibly a right
child $t1$.  The leaf positions of the transaction tree
are always populated left to right, heap-style.
In \reffig{f:domainsplit}, the transaction tree is implicit in the
containment relation of boxes: the outermost box is the root,
which contains group 0 and 1; group 0 contains transactions 00,01
and group 1 contains transactions 10,11.

Each transaction node maintains a count of leafs in its subtree,
so we can quickly determine whether a subtree can incorporate another
incoming transaction (\refsec{s:intake}).

%%%%%%%%%%%%%%%%%%%%%%%%%%%%%%%%%%%%%%%%%%%%%%%%%%%%%%%%%%%%%%%%%%%%%%%%%
\subsection{Signals}

\label{s:signals}

A \emph{signal} is an information flow between operators.
Signals are realized by data structures; between machines,
changes to signals are communicated by protocol.
Signal names bear sub- and superscript labels indicating (respectively)
their transaction group and subdomain.  For example, the signal
$\delta_{00}^{010}$ carries deltas from the first (00) transaction (group)
that occur in the third (010) subdomain (the first two being, of course,
000 and 001.)
Signals are emitted by operators (\refsec{s:operators}); each time an operator
is refreshed, it emits updates to all of its output signals
in a single atomic operation.

{\bf Delta signals} ($\delta_t^d$) carry changes to database state.
Each delta record is of the form
$(A,\overline{k},\overline{v},\pm)$, where $A$ identifies a predicate,
$\overline{k}$ is a key tuple, $\overline{v}$ is a value tuple, and
$\pm \in \{ +,- \}$ distinguishes \emph{upserts} and \emph{retractions}.
An \emph{upsert} $(A,\overline{k},\overline{v},+)$
indicates the update or insertion of a record
(hence the portmanteau).  For a relation $R(\overline{k})$, the
value tuple $\overline{v}$ is omitted
(or construed to be the empty tuple).  For a scalar quantity
(i.e. a nullary function), $\overline{k}$ is an empty tuple.
A retraction $(R,\overline{k},\cdot,-)$
indicates removal of a record; the value tuple is omitted.

{\bf Correction signals} ($c_t^d$) also carry changes to database state.
They have the same format as delta signals; the name
\emph{correction} distinguishes them as signals carrying relevant
deltas back toward transactions in need of repair.
(When a delta signal is filtered with sensitivities by a
$\mathsf{corr}$ operator, it becomes a correction signal.)

{\bf Sensitivity signals} ($s_t^d$) carry information about what aspects
of database state might require repairing a transaction, if changed.
Each sensitivity record is of the form $(A,[\overline{k}_1,\overline{k}_2])$,
where $A$ identifies a predicate, and
$[\overline{k}_1,\overline{k}_2]$ is an interval of key tuples.
For efficiency, sensitivity records can be represented in the
form $(A,\overline{k}_0,[\overline{k}_1',\overline{k}_2'])$, where
$\overline{k}_0$ is a prefix common to $\overline{k}_1$ and
$\overline{k}_2$, and $\overline{k}_1',\overline{k}_2'$ give
the remaining tuple elements.  This form of sensitivity intervals
is emitted by the incremental variant of leapfrog triejoin
(\refsec{s:lftj}).

\subsubsection{Signals associated with transaction nodes}

Leaf nodes of the transaction tree have \emph{height} 0; a group
node has height one greater than its children.  A tree node (leaf or
group) of height $h$ is associated with the following signals:
\begin{itemize}
\item $2^h$ delta signals, named $\delta_t^{0\cdots000}$, 
$\delta_t^{0\cdots001}$, $\delta_t^{0\cdots010}$ etc., where the domain
labels have $h$ binary digits.  (Leaf nodes have empty domain labels.)
\item $2^h$ sensitivity signals $s_t^{0\cdots00}$, etc.
\item $2^h$ correction signals $c_t^{0\cdots00}$, etc.
\end{itemize}

%%%%%%%%%%%%%%%%%%%%%%%%%%%%%%%%%%%%%%%%%%%%%%%%%%%%%%%%%%%%%%%%%%%%%%%%%
\subsubsection{Data structures for signals}

A signal can undergo multiple revisions as transactions are repaired.
To avoid concurrency hazards, one can either employ read-write locks
(i.e. each operator acquires read locks on signals it inputs, and
write locks on signals it outputs, and is postponed until such
locks can be acquired); or one can use a lock-free, purely functional
representation.  We have used both in practice, and have a
slight preference for purely functional data structures, as they
simpler, intrinsically scalable, and adapt easily to repair circuits
that span multiple machines.

For efficiency, we want to incrementally maintain operators
when changes occur to their signals, rather than evaluating from scratch
each time an operator is \emph{refreshed}.  For this reason, signals
must support versioning and efficient enumeration of changes between
versions: each time a signal is read, it has an associated version-id
(an ordinal), and we must be able to enumerate changes between two
specified version-ids in time $O(\Delta \log n)$ where $n$ is
the number of records in the latest version, and $\Delta$ is
the number of update operations (i.e. records inserted or removed) performed
between the specified versions.  This can be achieved by a variety of
purely functional, versioned, or persistent data structures
\cite{Driscoll:STOC:1986,Okasaki:1998}.
We employ data structures with efficient iterators that support seeking
i.e. $O(m \cdot \log(n/m))$ time to visit $m$ of $n$ records, in order.

%%%%%%%%%%%%%%%%%%%%%%%%%%%%%%%%%%%%%%%%%%%%%%%%%%%%%%%%%%%%%%%%%%%%%%%%%
\subsection{Operators}

\label{s:operators}
Signals are processed by \emph{operators}: each operator inputs
a few signals, and outputs one or two signals.
An operator tracks the version-id for its inputs, so that each
time an operator is refreshed it can efficiently enumerate
changes that have occurred to its inputs.

Three operators are employed: $\mathsf{txn}$, $\mathsf{merge}$,
and $\mathsf{corr}$.  For each, we describe their
operation in both full evaluation and incremental maintenance mode.

%%%%%%%%%%%%%%%%%%%%%%%%%%%%%%%%%%%%%%%%%%%%%%%%%%%%%%%%%%%%%%%%%%%%%%%%%
\subsubsection{The txn operator}

\label{s:txnop}

Each transaction $t$ is implemented by an operator $\mathsf{txn}_t$
which reads the
database state and corrections $c_t$ that reflect relevant updates
to the database state by older transactions.  A transaction outputs
a delta signal $\delta_t$ which indicates changes it makes to the
database state, and a sensitivities signal $s_t$ indicating
aspects of the database state it read or attempted to read:
\begin{align}
\mathsf{txn}_t \left(
\mathsf{db}, c_t \right) &\Rightarrow (s_{t}, \delta_{t})
\end{align}

We describe the details of individual rules and transactions
in Sections \ref{s:lftj} and \ref{s:graph}.  In summary,
a transaction consists of multiple rules that may
query database state and request updates of it.  In full evaluation
mode, each transaction rule is evaluated.  Rule evaluation
emits sensitivity records, originating from incremental
leapfrog triejoin (\ref{s:lftj}).  Some rules may produce
deltas (requested changes to database state).
In incremental evaluation mode, the
transaction is repaired for relevant corrections made to the database
state, as carried by the correction signal $c_t$ (\refsec{s:graph}).

If an error occurs in a transaction that would require its
abort in serial execution (e.g. an integrity constraint
failure), the output signals are revised
to be empty: $\delta_t = \emptyset$, $s_t = \emptyset$.
The transaction is not officially aborted until all its input
signals have been marked as finalized (\refsec{s:finalization});
a failure might be transient, and the transaction
might succeed once repaired for changes made by previous transactions.

%%%%%%%%%%%%%%%%%%%%%%%%%%%%%%%%%%%%%%%%%%%%%%%%%%%%%%%%%%%%%%%%%%%%%%%%%
\subsubsection{The merge operator}

\label{s:merge}

The $\mathsf{merge}$ operator takes a pair of transaction-split
signals and produces a pair of domain-split signals.  Merge operators
are used to combine delta signals, and to combine sensitivity signals
(\reffig{f:domainsplit}).

For sensitivity signals, a use of the merge operator has the form:
\begin{align}
\mathsf{merge} \left( 
 \left[
 \begin{array}{c}
   s_{t0}^d \\
   s_{t1}^d
 \end{array}
 \right]
 \right)
&\Rightarrow
 \left[
 \begin{array}{c}
   s_t^{d0} \\
   s_t^{d1}
 \end{array}
 \right]
\end{align}

When a merge operator is constructed, it is given a
\emph{split point} of the form $(R,\overline{k})$, where
$R$ identifies a database relation, and $\overline{k}$ is
a key tuple (\refsec{s:domaindecomp}).
Each sensitivity record represents an interval (\refsec{s:signals}).
Records whose key interval endpoint is
$< \overline{k}$ are placed in $s_t^{d0}$, and those whose interval
start is $\geq \overline{k}$ are placed in $s_t^{d1}$.  If a sensitivity
record $(R,[\overline{k}_1,\overline{k}_2])$ straddles the split point,
two records are emitted: $(R,[\overline{k}_1,\overline{k}])$ is placed in 
$s_t^{d0}$, and $(R,[\overline{k},\overline{k}_2])$ is placed in
$s_t^{d1}$.  Sensitivity records from the two transaction
subtrees may overlap; one can eliminate duplicates,
merge overlapping intervals, and so forth.

The merge operator for delta signals is simpler.
Delta signals carry records of the
form $(R,\overline{k}',\overline{v},\pm)$; records with $\overline{k}' < \overline{k}$ (below the split point)
are placed in $\delta_t^{d0}$, and those with $\overline{k}' \geq \overline{k}$ are placed
in $\delta_t^{d1}$.
When a record for a key $\overline{k}'$ occurs in both $\delta_{t0}^d$ and
$\delta_{t1}^d$, the delta from $\delta_{t1}^d$ is given priority,
since it originated from a later transaction.  Note that when
$\delta_{t0}^d$ contains an upsert record $(R,\overline{k},\overline{v},+)$
and $\delta_{t1}^d$ contains a retraction record $(R,\overline{k}',\cdot,-)$,
the retraction record is emitted; the retraction and insertion do not
annihilate.

Full evaluation: using a pair of iterators for the incoming signals,
one can iterate the records in order, and write corresponding record(s) to
the outgoing subdomain signals as described above.
If an identical
record occurs in both input signals, only one instance 
need be emitted.

Incremental evaluation: for each changed record, revisit the merge for
that record, revising the output signal appropriately.

%%%%%%%%%%%%%%%%%%%%%%%%%%%%%%%%%%%%%%%%%%%%%%%%%%%%%%%%%%%%%%%%%%%%%%%%%
\subsubsection{The corr operator}

\label{s:corr}

Conceptually, the $\mathsf{corr}$ filter is responsible for
intersecting a sensitivity signal with delta and correction signals,
propagating relevant corrections toward transactions in need of repair.
\begin{align}
\mathsf{corr} \left( s_{t0}^d,
[ c_{t}^{d0}, c_{t}^{d1} ], \emptyset \right) & \Rightarrow c^d_{t0} \\
\mathsf{corr} \left( s_{t1}^d,
[ c_{t}^{d0}, c_{t}^{d1} ], \delta_t^{d0} \right) & \Rightarrow c^d_{t1}
\end{align}

Each $\mathsf{corr}$ operator takes a sensitivity signal,
a pair of domain-split correction signals, and an optional
delta signal.  (The delta signal is used for right children
of transaction nodes, but not for left children.)

In full evaluation mode, one can choose among multiple strategies,
basing the decision on whether there are many sensitivities and
few corrections/deltas, or vice versa; one can choose a
different strategy for each predicate P mentioned in the
signals.  In all cases, records from the delta signal
supersede those of the correction signal,
since they originate from later transactions.
\begin{enumerate}
\item If there are comparatively few sensitivity records,
one can iterate the sensitivity records of $s^d_{t1}$.  For each
sensitivity record, extract from $\delta_t^{d0}, c_{t}^{d0}, c_{t}^{d1}$
all record(s) which intersect.
\item If there are many sensitivity records, and comparably
few corrections and deltas, one can iterate the corrections
and deltas, and for each record look for a sensitivity record
with a containing interval.  Special data structure support
is required, as is done for incremental leapfrog triejoin
(\refsec{s:inclftj}).  Only one matching sensitivity interval
is required; one need not find all matching intervals, as
in incremental leapfrog triejoin.
\end{enumerate}

Incremental evaluation:
\begin{enumerate}
\item If the sensitivity signal is unchanged since the last refresh:
For each of $\delta_t^{d0}, c_{t}^{d0}, c_{t}^{d1}$, enumerate
the changes since the last refresh, collecting key tuples
in a priority queue.
Iterate through these keys in order.  For each key, determine the
net change (for example, an record might be removed from $c_t^{d0}$
but also inserted to $\delta_t^{d0}$, a net insertion.\footnote{
This can be somewhat more confusing that it first sounds.
The delta record $(A,100,-)$ indicates the removal of
$A(100)$.  When iterating \emph{changes} to the delta signal,
one might encounter a change $((A,100,-),-)$, indicating that
the record recording the removal of $A(100)$ was itself removed.
When matching sensitivity intervals, we are not interested in
the inner $\pm$, only whether a key for a delta record (either
an insert or erase) matches an interval.}
For removed records, remove them from the output signal, if present.
For inserted records, search for a matching sensitivity interval,
as in strategy(2) above.
\item If the sensitivity signal has changed, but the
correction and delta signals have not, iterate changes to the
sensitivity signal, and use iterators for
$\delta_t^{d0}, c_{t}^{d0}, c_{t}^{d1}$ to efficiently seek
matching records.  For inserted sensitivity intervals, matching
records are placed in the outgoingcorrection signal.  For
removed sensitivity intervals, one
must determine whether a matching sensitivity interval still exists 
for each correction/delta record contained in the removed interval.
(Alternately, one can maintain a count of the number of matching
sensitivity intervals for each correction/delta, and remove a
correction/delta when its count reaches zero.)
\item If both the sensitivity signal and one or more of the
delta \& sensitivity signals have changed, apply both (1) and
(2), using the previous version of the sensitivity signal
in (1), and then the newer version of all signals in (2).
\end{enumerate}

%%%%%%%%%%%%%%%%%%%%%%%%%%%%%%%%%%%%%%%%%%%%%%%%%%%%%%%%%%%%%%%%%%%%%%%%%
\subsection{Summary of wiring}

The following equations summarize the operators and signals
of a repair circuit.
\begin{align}
\mathsf{txn}_t \left(
\mathsf{db}, c_t \right) &\Rightarrow (s_{t}, \delta_{t}) \label{e:txn} \\
\mathsf{merge} \left(
 \left[ 
 \begin{array}{c}
   \delta_{t0}^d \\
   \delta_{t1}^d
 \end{array}
 \right]
 \right) & \Rightarrow
 \left[
  \begin{array}{c}
   \delta_t^{d0} \\
   \delta_t^{d1} 
  \end{array} \right] \\
\mathsf{merge} \left( 
 \left[
 \begin{array}{c}
   s_{t0}^d \\
   s_{t1}^d
 \end{array}
 \right]
 \right)
&\Rightarrow
 \left[
 \begin{array}{c}
   s_t^{d0} \\
   s_t^{d1}
 \end{array}
 \right] \\
\mathsf{corr} \left( s_{t0}^d,
[ c_{t}^{d0}, c_{t}^{d1} ], \emptyset \right) & \Rightarrow c^d_{t0} \\
\mathsf{corr} \left( s_{t1}^d,
[ c_{t}^{d0}, c_{t}^{d1} ], \delta_t^{d0} \right) & \Rightarrow c^d_{t1}
\end{align}

\begin{enumerate}
\item For each transaction leaf node, we instantiate the transaction
operator \refeqn{e:txn}.
\item At each transaction tree group node $t$ of height $h$, we
instantiate $\mathsf{merge}$ and $\mathsf{corr}$ equations for all
$2^h$ binary strings $d$ of length $h$, e.g., for height h=3 we would use
$d \in \{ 000$, $001$, $010$, $\ldots$, $111 \}$.
\end{enumerate}

\section{Transaction repair in action}
\label{s:running}

%%%%%%%%%%%%%%%%%%%%%%%%%%%%%%%%%%%%%%%%%%%%%%%%%%%%%%%%%%%%%%%%%%%%%%%%%
\subsection{Fixpoint mechanics}

\label{s:priorities}

Conceptually, the transaction repair circuit is evaluated using a fixpoint
iteration.  A naive iteration would initially set all signals
(apart from initial database state) to $\emptyset$.
In the first iteration one would evaluate
every operator; in subsequent iterations, refreshing only operators
whose input signals changed in the previous iteration.

Time can be saved by using an appropriate
\emph{relaxation} of the iteration schedule.
In a relaxed schedule, we do not refresh operators in a lockstep
global fixpoint iteration, but rather refresh individual operators
opportunistically and in parallel with no fixed schedule; each
time a compute unit becomes available, we select an operator for
it to refresh.  In addition, we require that each operator be
selected for refresh at least once, and an operator with a changed input
must eventually be refreshed.  This style of fixpoint evaluation is known as
an \emph{asynchronous} (or \emph{chaotic}) relaxation.

%%%%%%%%%%%%%%%%%%%%%%%%%%%%%%%%%%%%%%%%%%%%%%%%%%%%%%%%%%%%%%%%%%%%%%%%%
\subsubsection{Soundness and convergence}

\label{s:convergence}

An obvious concern is whether asynchronous relaxation is sound
and convergent for transaction repair.  We make an informal 
convergence argument by induction over the serialization order.
The convergence is to a unique fixpoint, hence sound.
The argument relies on some assumptions:
\begin{enumerate}
\item The database is finite.
\item Operators are purely functional, i.e., their output signals
are uniquely determined by their input signals.
\item The evaluation (full or incremental) of an operator
always terminates.  (For individual transactions, this is enforced by the
PTIME bound of our language.)
\item We force the outgoing sensitivity signals of individual transactions
to be monotone by never removing records from them.
A transaction can contain rules that would otherwise result in nonmonotone
updates to the sensitivity signal.  For example, a transaction could examine
a record $E(10)$ only when a record $D(10)$ is absent; a
correction signal bringing the news that $D(10)$ is present
would result in the sensitivities decreasing.  This makes it
challenging if not impossible to fashion a convergence argument.
We therefore require the sensitivity signal from a transaction
to be increasing.  Since sensitivity signals from
individual transactions are monotone, all sensitivity signals
in the circuit are, since they are collations of single-transaction
sensitivities.
\item We assume individual operators correctly implement their 
functionality.  For example, if we iterate a single $\mathsf{txn}$ 
operator until it reaches a fixed point (holding the incoming corrections
constant), then no subsequent changes made to the incoming corrections
outside its reported sensitivity regions will affect it.
\end{enumerate}

We proceed by induction over the serialization order.
For the base case: since transaction 0 depends only on the
initial database state, which is fixed for the course of the
iteration, it will reach its final state the first time it is
refreshed.

Induction step: 
by design, information carried by correction and delta signals flows 
strictly forward, following the serialization order: no information
flows from transaction $j$ to transaction $i$ when $i < j$.
If all transactions $< i$ have
converged, then delta, sensitivity and correction signals depending
solely on transactions $< i$ will converge, as every cycle of such
signals is interrupted by a transaction $< i$, whose output signals
have converged.

We encounter a slight wrinkle in arguing that
transaction $i+1$ will converge.  This transaction receives
a correction signal from transactions $< i$, and possibly
a delta signal from transaction $i$; these signals are modulated
by the sensitivity signal emitted by transaction $i+1$ itself
(\reffig{f:twotxns-c}, with the top transaction mentally
labelled $i$).  Here we invoke the enforced monotonicity of the
sensitivity signal and finiteness of the database: the sensitivity
signal for transaction $i+1$ must eventually converge, since its
lattice is finite.  Once its sensitivity signal convergences, the
convergence of transaction $i+1$ follows by the purely functional
operator property.

%%%%%%%%%%%%%%%%%%%%%%%%%%%%%%%%%%%%%%%%%%%%%%%%%%%%%%%%%%%%%%%%%%%%%%%%%
\subsubsection{Priorities}

In asynchronous relaxation, it is useful to define a priority on operators.
When a compute unit becomes available, we select the operator of
highest priority in need of refresh.  

The importance of selecting
priorities wisely is best illustrated by a bad choice: suppose
we gave transactions latest in the serialization highest priority.
If we had transactions numbered $0,\ldots,k-1$, each incrementing
the same scalar counter, the order of transaction refreshes
would be $k-1$, then $k-2,k-1$, then $k-3,k-2,k-1$, etc.
i.e. $\Theta(k^2)$ transaction repairs.
Similarly bad things happen if operators at height $h$ of
the transaction tree are given priority over those at height $h-1$. 

We assign priorities
according to the following principles:
\begin{enumerate}
\item Transactions are given priorities reflecting the serialization order,
with the earliest transaction given highest priority.
\item We label each signal with the priority of the operator generating it.
We then assign priority to operators to be lower than their inputs.
This encourages operators to wait until all their inputs have
converged.
Since every signal-cycle contains a transaction, and transactions
are assigned fixed priorities, this definition of priorities
is well-founded.
\item In (2), we select priorities so that signals depending solely
on transactions $< i$ have greater priority than transaction $i$.
\end{enumerate}

%%%%%%%%%%%%%%%%%%%%%%%%%%%%%%%%%%%%%%%%%%%%%%%%%%%%%%%%%%%%%%%%%%%%%%%%%
\subsubsection{Enqueuings}

We use a priority queue to track operators in need of refresh.
Operators are placed in the queue when these events occur:
\begin{enumerate}
\item When a new transaction is added to the transaction tree, its
$\mathsf{txn}$ operator is enqueued.
\item When an operator is refreshed, it may change some of its output signals.
For each changed output signal, all operators reading that signal
are enqueued (if they are not already).
\end{enumerate}

%%%%%%%%%%%%%%%%%%%%%%%%%%%%%%%%%%%%%%%%%%%%%%%%%%%%%%%%%%%%%%%%%%%%%%%%%
\subsubsection{Finalization of signals and transactions}

\label{s:finalization}

We say a transaction is \emph{finalized} when it has converged
in the sense of \refsec{s:convergence}.
Tracking which transactions have been finalized plays a
central role in the commit mechanics (\refsec{s:commit}).
The first transaction in the serialization order is finalized
at completion of its first evaluation; the first transaction
never needs to be repaired.  A signal is finalized if it is
emitted by a finalized operator; an operator is finalized if
all its inputs are finalized.  Individual transactions are
a special case, due to the cycles between transactions,
sensitivities, and corrections.
A transaction is finalized once
(a) all previous transactions in the serialization order
are finalized; and (b) no signals participating in a cycle
with the transaction are downstream from an operator either
being refreshed or in the queue awaiting refresh.

Once a transaction is finalized, we notify its originator
that the transaction has been accepted.  (The LogicBlox database
sends two notifications for transactions: first of acceptance,
and then of durable commit and/or replication.)

%%%%%%%%%%%%%%%%%%%%%%%%%%%%%%%%%%%%%%%%%%%%%%%%%%%%%%%%%%%%%%%%%%%%%%%%%
\subsection{Intake and commit of transactions}

\label{s:intake}

In practice the transaction repair circuit is being grown
and pruned dynamically as new transactions arrive, and
finalized transactions are committed.
For efficiency, we try to keep the transaction tree small,
since there are additional costs with each
increase in height of the tree.  One way we do this is by
pruning transactions after commit (\refsec{s:commit}).
Another is by deferring intake of new transactions.
If transactions are balanced in load, then we only
need roughly as many transactions in the tree as there are
compute units, plus a number of transactions whose
processing time is equivalent to the commit pipeline
latency.

When new transactions arrive, they are not immediately
added to the transaction tree.  Instead they are placed
in a holding queue.  While there, they may be triaged
and their placement in the serialization order optimized
(\refsec{s:arrangement}).

When a compute unit seeking an operator to refresh finds
none eligible, it selects a transaction from the holding
queue and moves it into the transaction tree.  Transactions
are always inserted to the leftmost unoccupied leaf position.
If the transaction tree is full, another layer of height is
added by creating a new root node whose left child is the
current root node (\reffig{f:growshrink}, left).

%%%%%%%%%%%%%%%%%%%%%%%%%%%%%%%%%%%%%%%%%%%%%%%%%%%%%%%%%%%%%%%%%%%%%%%%%
\subsubsection{Null transactions}

\emph{Null transactions} can be employed for load balancing, and
to accommodate configurations that do not have a number of machines equal
to a power of two.  Null transactions produce no deltas or
sensitivities; their presence in a circuit does not substantially
affect performance, if employed thoughtfully.  If transactions require
uneven amounts of effort, null transactions can be inserted into
the serialization order to alter the allocation of transactions.
For arbitrary number of machines, one can pad up to the nearest
power of two by inserting phantom machines processing
null transactions.  (This requires adjustments to domain
decomposition to ensure load balance.)

%%%%%%%%%%%%%%%%%%%%%%%%%%%%%%%%%%%%%%%%%%%%%%%%%%%%%%%%%%%%%%%%%%%%%%%%%
\subsubsection{Handling long-running transactions}

Transactions do not report any deltas or sensitivities until
they complete their initial evaluation.  This means we can
bump long running transaction(s) from a potential commit
set and replace them with null transactions.

We can in general bump arbitrary transactions, even those
that have already reported deltas and are undergoing repair.
We simply replace the transaction with a null transaction
(whose empty delta and sensitivity sets will then propagate
out, removing the deltas from the bumped transaction).
We can then place the transaction in a new position.
To avoid discarding all the repair done so far, we can
put the transaction on hold until its sensitivities have
propagated out to the root level of the transaction
tree, and adjusted corrections have flowed back.

%%%%%%%%%%%%%%%%%%%%%%%%%%%%%%%%%%%%%%%%%%%%%%%%%%%%%%%%%%%%%%%%%%%%%%%%%
\subsubsection{Commit mechanics}

\label{s:commit}

\begin{figure}[t]
\includegraphics[width=\columnwidth]{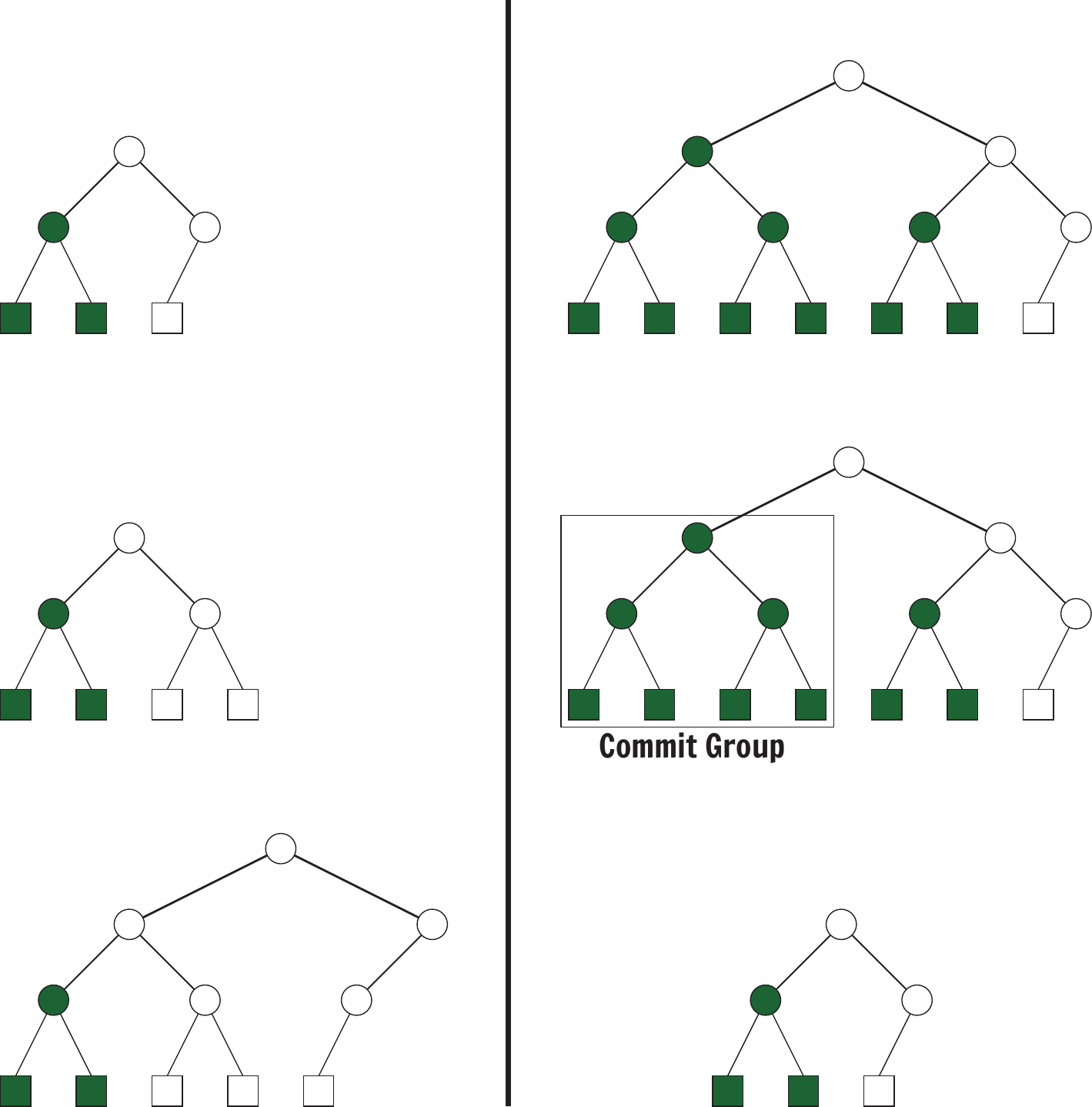}
\caption{\label{f:growshrink}Growing and shrinking the transaction
tree.  Left, top to bottom: new transactions are placed in the
first available leaf position.  When the transaction tree becomes
full, a new root node is added, with the current tree becoming
its left subtree.  Right, top to bottom: finalized transactions
(and groups) are marked green.  The simplest style of commit chooses
a finalized left child of the root node (middle).  When the left child
of the root has been committed, the right child becomes the
new transaction tree (bottom).}
\end{figure}

\begin{figure}[t]
\includegraphics[width=\columnwidth]{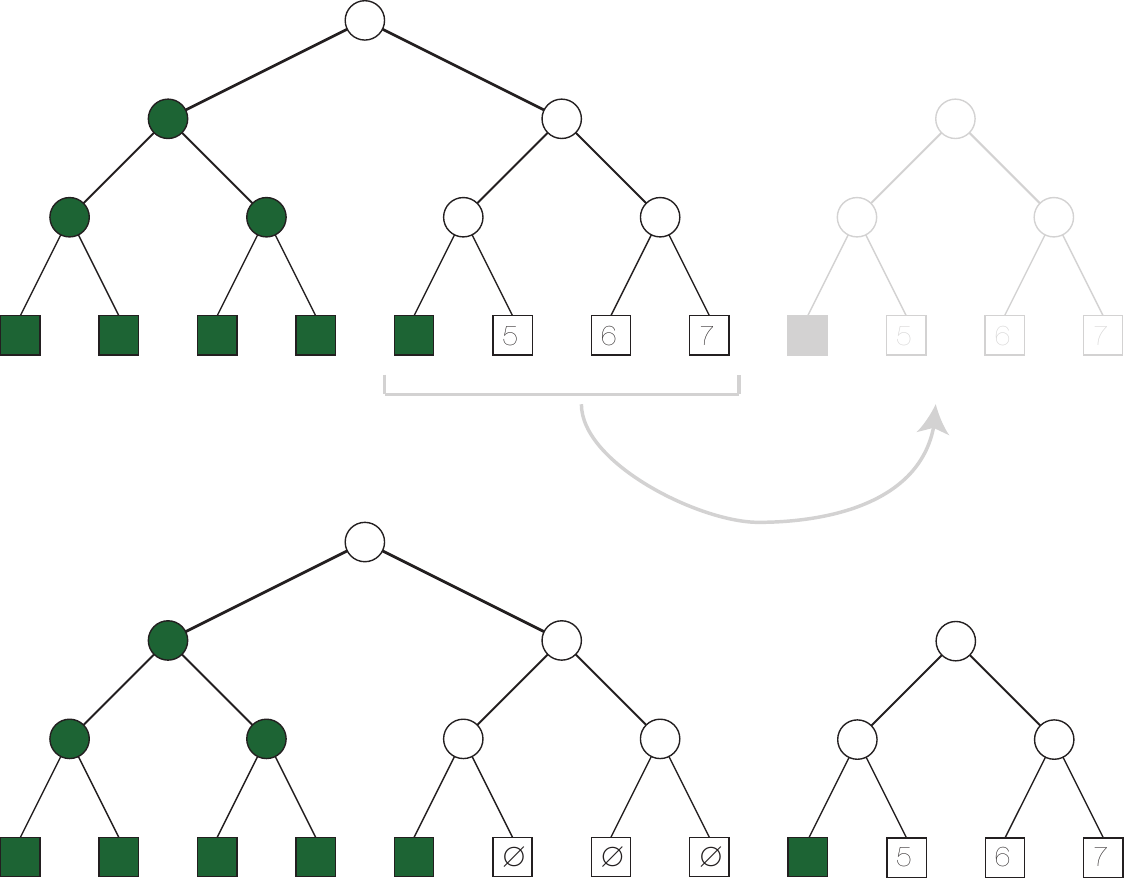}
\caption{\label{f:commitpadding}Commit padding.  A more ambitious
commit can tackle the entire sequence of finalized transactions.
The smallest subtree containing all non-finalized transactions
becomes the new transaction tree, with finalized transactions
replaced by null transactions.  Non-finalized transactions in
the current transaction tree are replaced with null transactions;
once the signals for this tree become finalized, the commit can
proceed with the apply-deltas phase.}
\end{figure}

Our durable commit process is arranged in a somewhat complex pipline,
the details of which revolve around particularities of our metadata
and page management techniques.  The relevant point is that
when the first stage of the commit pipeline is idle, it grabs
a new batch of transactions and starts them down the pipeline.
This batch of transactions is committed as a group, with all
members of the group being notified simultaneously when the
durable commit and/or replications complete.

There are two strategies that can be employed.  The simplest can
be used when the root node's left subtree contains only finalized
transactions.  In this case we can select the transactions of
the left subtree for commit, and place them into the commit pipeline.
This strategy is advantageous because the
delta signals of the left subtree contain exactly the deltas
to be applied to the database state.

Whenever the root node
of the transaction tree has a left child whose subtree contains
only transactions which have been submitted to the commit pipeline,
we discard the left subtree, and the right child of the root node
becomes the new root node (\reffig{f:growshrink}, right).
This discourages the transaction tree
from becoming arbitrarily high.
(Note though, that signals from the discarded subtree may
remain in use by transactions in the tree; they are not
garbage collected until they become inaccessible.)

A slightly more complicated strategy can be used to commit
all transactions which have been finalized.  This involves
duplicating the transaction tree, an $O(1)$ operation if
purely functional data structures are used.
In one version we replace
all non-finalized transactions with null transactions, and
wait for its top-level delta signals to converge
before submitting to the commit pipeline; this ensures no
deltas are included from non-finalized transactions.
In the other version of the transaction tree,
we prune to the smallest subtree containing
all nonfinalized transactions (\reffig{f:commitpadding}).

\begin{figure}[t]
\includegraphics[width=\columnwidth]{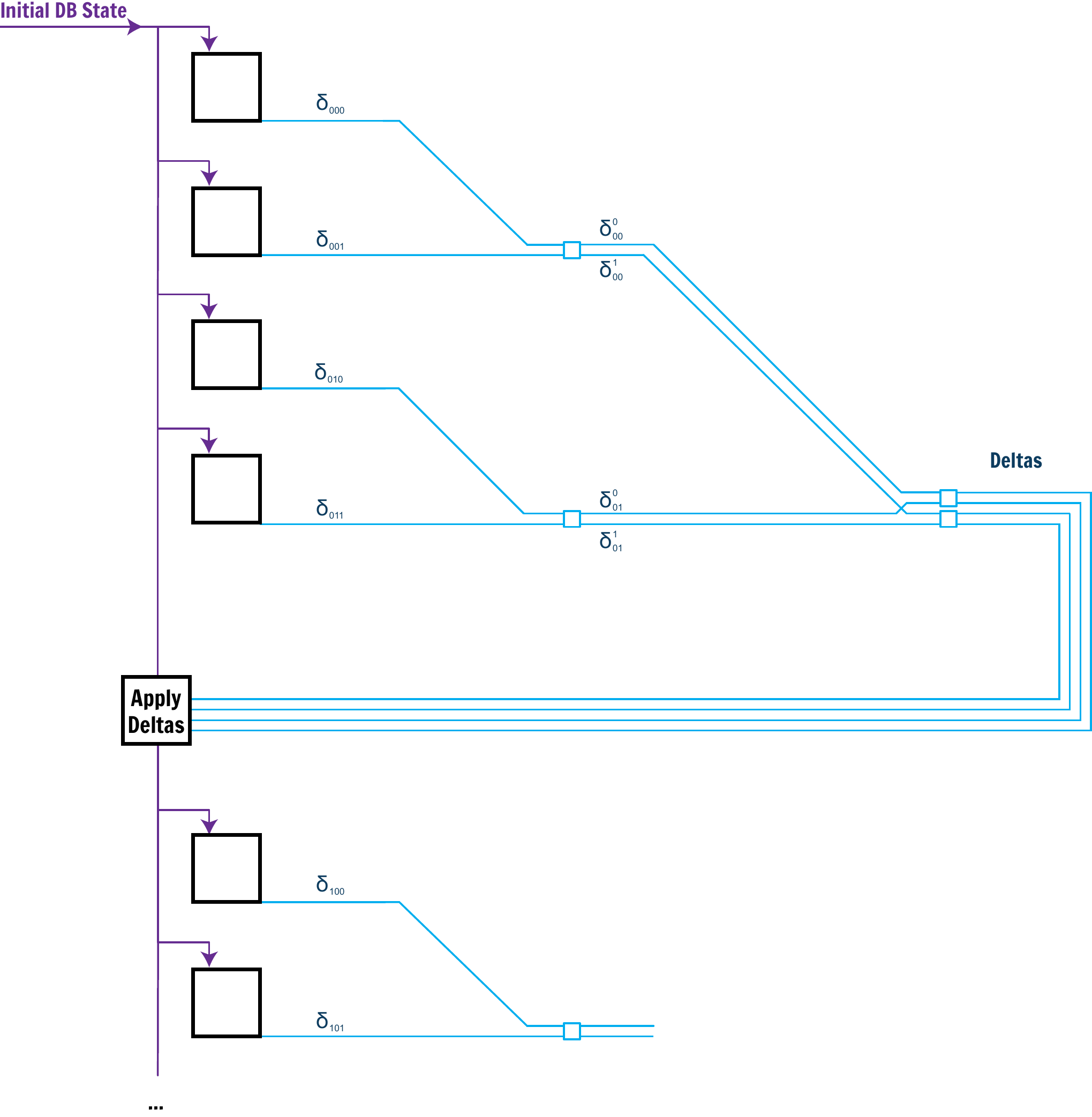}
\end{figure}

As the first step in committing a group of finalized transactions,
we apply their net deltas to the database state.  This can be
accomplished by submitting high-priority tasks to the same pool
of workers used to refresh operators, with one task for each
of the $2^h$ subdomains.  Each task reads a delta signal
for a subdomain, and applies the changes to a branch of the
database state.  Once all these tasks finish, the branch becomes
the new tip version of the database.  Transactions arriving after
this event are given this new version as their initial database state,
and corrections only from the transactions remaining in the tree.

\section{Leapfrog Triejoin}

\label{s:lftj}

In previous sections we have described transaction repair at
a high level of detail, with a transaction presented as a
black box.  In this section and \refsec{s:graph} we open
the box, detailing repair mechanisms at the level of
individual transactions.

The LogicBlox database processes transactions that
consist of one or more \emph{rules} written in our LogiQL
language, a descendent of Datalog.  A simple example of
a rule:
\begin{align}
D(x,y) &\leftarrow A(x), B(x,y), C(y).
\label{e:ABC}
\end{align}
To the left of the $\leftarrow$ is the rule head
listing inferences to be made when satisfying assignments
are found for the rule body $A(x), B(x,y), C(y)$.
Most rules have straightforward interpretations as sentences
of first-order logic:
\begin{align}
\forall x,y,z ~.~ \left[ A(x) \wedge B(x,y) \wedge C(y) \right] \rightarrow D(x,y).
\end{align}

\noindent
The LogiQL language has a rich feature set, but most functionality
(in particular, all rule bodies) can be lowered to a minimal core
language used by our optimizer and rule evaluators.  The syntax of
the core language is described by this grammar:
\begin{align*}
\mathsf{rule} &::= \forall \overline{x} ~.~ \mathsf{head} \leftarrow \mathsf{conj} \\
\mathsf{head} &::= \mathsf{atom} ~(~ , \mathsf{atom} )^\ast \\
\mathsf{atom} &::= R(\overline{y}) ~|~ F[\overline{y}]=\overline{z} \\
\mathsf{conj} &::= ~[~ \exists \overline{x} ~.~ ~]~ \mathsf{dform} ~(~, \mathsf{dform} ~)^\ast
~ \\
\mathsf{dform} &::= \mathsf{atom} ~|~ \mathsf{disj} ~|~ \mathsf{negation} \\
\mathsf{disj} &::= \mathsf{conj} ~(~ ; \mathsf{conj} )^+ \\
\mathsf{negation} &::= \mbox{!} ( \mathsf{conj})
\end{align*}
Atoms can be either relations or functions, and may represent
either concrete data structures (representing edb functions/relations,
or materialized views),
or primitives such as addition and multiplication.
Rule bodies must usually adhere to existential ($\exists_1$) form, that is,
no existential quantifiers under an odd number of negations.\footnote{
Our aim is to avoid search under negation; we make exceptions for trivial
uses of existential quantifiers under negation, for example:
\begin{align*}
R(x) &\leftarrow A(x),!(\exists y ~.~ F[x]=y,~ y*y > 16)
\end{align*}
}
User rules not adhering to this form are rewritten by introducing
temporary predicates.   Sets of rules capture FO+lfp 
(i.e. PTIME).\footnote{For the brave and reckless, FO+PFP (PSPACE) if unsafe recursion
warnings are disabled.}

We use \emph{leapfrog triejoin} to enumerate satisfying assignments
of rule bodies.  Leapfrog triejoin is an algorithm for existential
queries that performs well over a variety of workloads,
and was recently proven to be worst-case optimal for full
conjunctive queries \cite{Veldhuizen:ICDT:2014}.
A few details are pertinent to transaction repair; for a more detailed
treatment we refer the reader to \cite{Veldhuizen:ICDT:2014}.

To evaluate a rule such as \refeqn{e:ABC}, our query optimizer selects an
efficient variable ordering for the instance; $[x,y]$, for example.
Leapfrog triejoin follows this variable ordering in performing a
backtracking search for satisfying assignments: it first
looks for $x$ which are present in both $A$ and $B(x,\_)$
(where $B(x,\_)$ is the projection $\{ x ~:~ \exists y . (x,y) \in B \}$ ).
For each such $x$, it searches for $y$ such that $(x,y) \in B$
and $y \in C$.  

For each variable a `leapfrogging' of iterators is performed,
selecting at each step an iterator and advancing it to a least-upper
bound for the positions of the other iterators.  \reffig{f:lftj1}
illustrates this process for $A = \{1,3,4,5,6,7\}$ and
$B(x,\_)=\{2,5,7\}$.  At $x=5$ all iterators come together on the same
value, and the search proceeds to the next variable, $y$ (\reffig{f:lftj1}, inset box).
When values of $y$ satisfying $B(5,y),C(y)$ are encountered, it emits the
current variable bindings as a satisfying assignment (e.g.
$x=5,y=101$).  When one
of the iterators reaches its end, the algorithm backtracks to
the previous variable, $x$, and continues the search.

\subsection{Incremental Leapfrog Triejoin}

\label{s:inclftj}

Leapfrog triejoin admits an efficient incremental evaluation algorithm
\cite{Veldhuizen:LB:2013}. 
For transaction repair this provides a
mechanism to quickly repair transaction rules
when corrections are made to the database state.
The LogicBlox database uses the same incremental evaluation 
algorithm for efficient
fixpoint computation and to maintain IDB predicates, that is,
materialized views installed in the database that are kept up-to-date
as transactions modify it.  

\begin{figure}[t]
\subfigure[\label{f:lftj1}Example of Leapfrog Triejoin in operation (see text).]{
\includegraphics[width=\columnwidth]{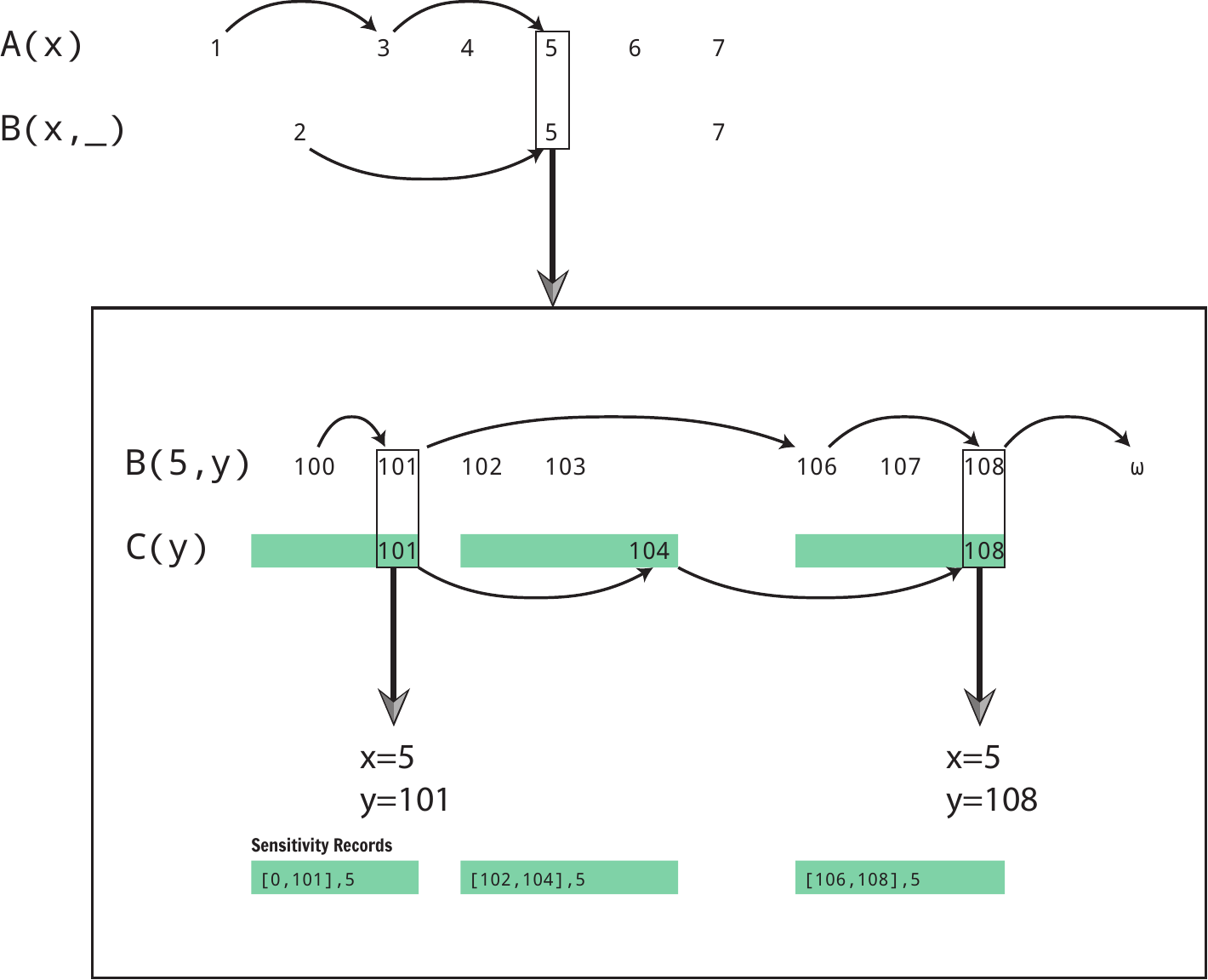}
} \\
\subfigure[\label{f:lftj2}Sensitivity records, some collected from iterator operations in \reffig{f:lftj1}, and a query for intervals matching $y=102$.]{
\includegraphics[width=\columnwidth]{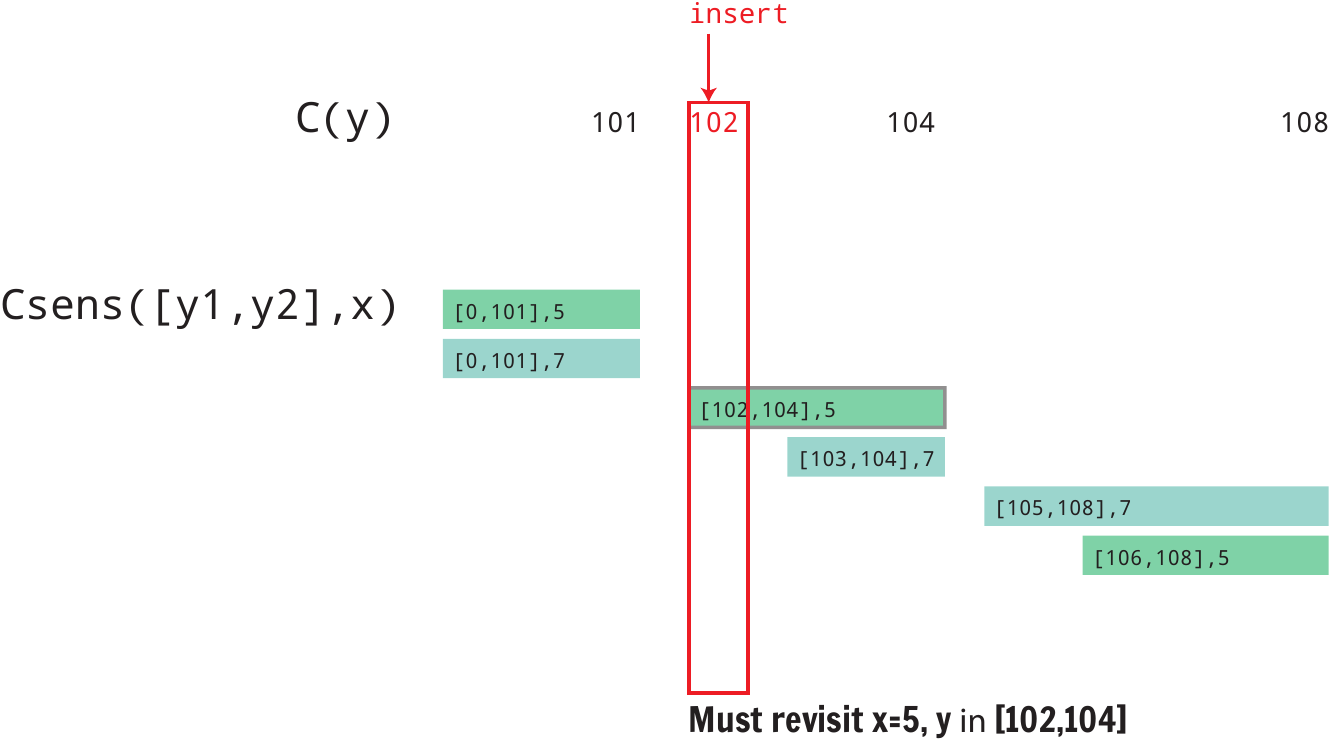}
}
\caption{Leapfrog Triejoin.}
\end{figure}

Our incremental evaluation algorithm for leapfrog triejoin
is closer to the trace-maintenance style
\cite{Acar:SODA:2004} than the approaches typically favoured
by the database community \cite{Chen:2005,Gupta:SIGMOD:1993,Gupta:1999}.
We can view \reffig{f:lftj1} as a trace
of the leapfrog triejoin algorithm, that is, a fine-grained
record of the steps performed in evaluating the rule.  By
recording a little information about the trace, we are able
to efficiently maintain the rule when one of the input
relations $A,B,C$ changes.

Consider the operation of the iterator for $C(y)$
in the inset box of \reffig{f:lftj1}.  The last
iterator operation for $C$ moves from 104 to 108; this
occurs because the iterator for B was positioned at
106, so the C iterator is asked to advance to a least upper
bound of 106.   How would the trace of this iterator operation
change if we removed or inserted elements to $C$?  Inserting 105 would
have no effect: the iterator would ignore 105
in seeking a least upper bound for 106.  Inserting
106 or 107 \emph{would} change the trace, as would removing 108.
In general, when an iterator is asked to seek a least upper
bound for a key $k$, it is sensitive to any change in the
interval $[k,k']$, where $k'$ is the current least upper
bound for $k$.  The coloured bars of \reffig{f:lftj1}
indicate these \emph{sensitivity intervals}.

We collect such intervals in \emph{sensitivity indices}.
For the $C(y)$ relation of this rule, the sensitivity index
has form $C_\mathit{sens}([y_1,y_2],x)$, where $[y_1,y_2]$ is a sensitivity
interval, and $x$ is a \emph{context key}.  For the iterator
operation that moves from 104 to 108, the sensitivity index
would collect a record $C_{\mathit{sens}}([106,108],5)$.
\reffig{f:lftj2} shows a sensitivity index for
which includes some sensitivity records collected during the
search for $y$ with $x=7$ (not shown).

\begin{figure}[t]
\includegraphics[width=\columnwidth]{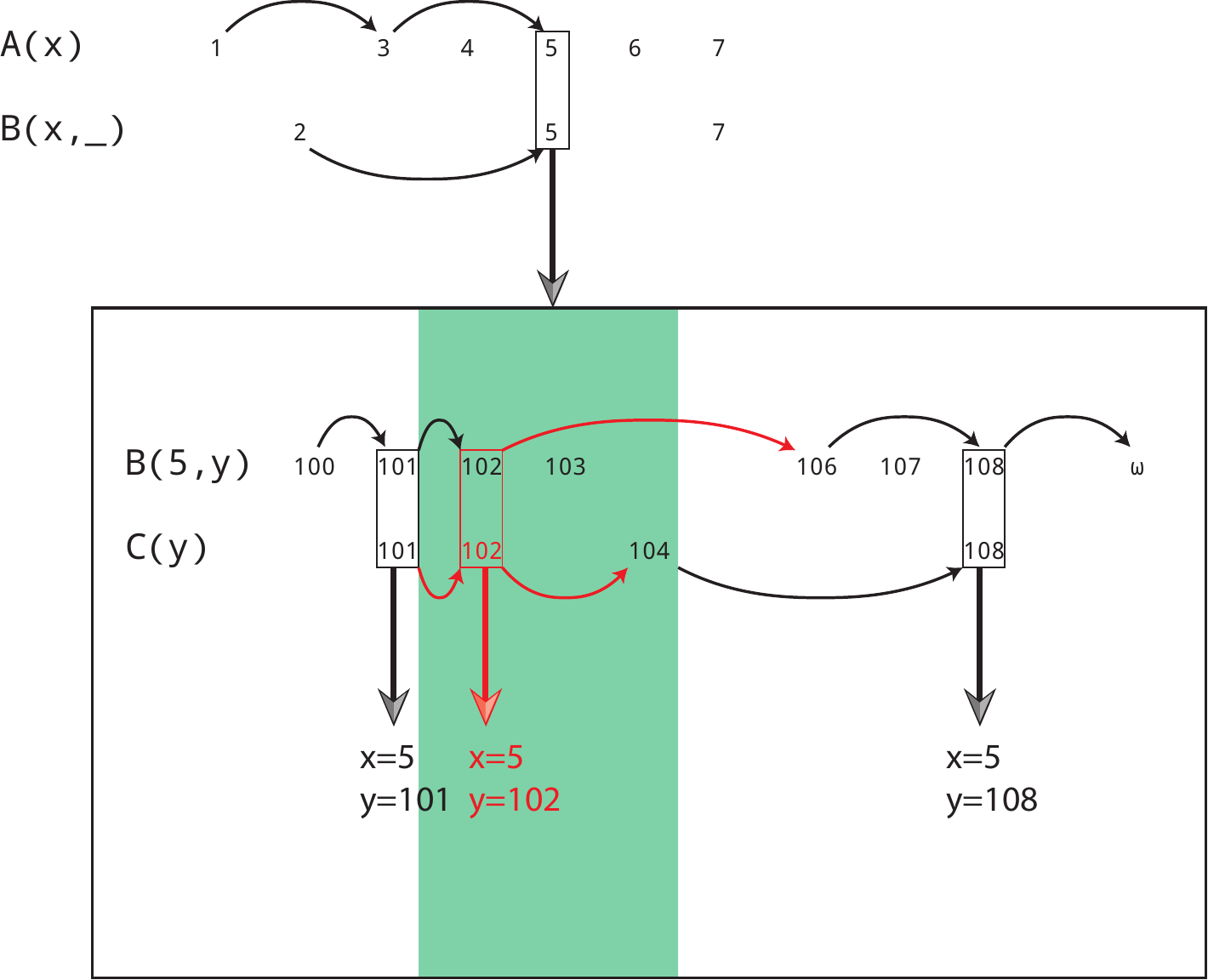}
\caption{\label{f:lftj3}The trace of \reffig{f:lftj1} after incremental
maintenance for insertion of $C(102)$.}
\end{figure}

Suppose we modify $C$
by inserting $102$, and we now wish to incrementally
maintain the rule.  To determine what regions of the trace
need revision, we query the sensitivity index to find intervals that
overlap the change at $y=102$.  We represent sensitivity indices
such as $C_\mathit{sens}([y_1,y_2],x)$ using a tree that permits
rapid computation of prefix sums \cite{Chatterjee:SUPER:1990};
its internal nodes are decorated with maximums of interval
endpoints $y_2$, allowing us to efficiently extract matching
intervals.  We find a match: $C_\mathit{sens}([102,104],5)$,
which suggests we revisit the trace where $x=5$ and
$102 \leq y \leq 104$.  Matching sensitivity intervals 
are used to construct the \emph{change oracle}, a nonmaterialized
view of the union of matching intervals.  We then evaluate a special
maintenance rule with form:
\begin{align}
\delta D(x,y,\Delta) &\leftarrow
\begin{array}[t]{l}
(\mathsf{Body}[A,B,C] \cdots \mathsf{Body}[A,B,C'])(x,y,\Delta), \\
\mathsf{ChangeOracle}(x,y).
\end{array}
\label{e:maintrule}
\end{align}
where $\delta D(x,y,\Delta)$ is the change in satisfying assignments
($\Delta \in \{+,-\}$),
$C'$ is the new version of the $C$ predicate,
\begin{align}
\mathsf{Body}[A,B,C](x,y) &\equiv A(x), B(x,y), C(y)
\end{align}
and $(\mathsf{Body}[A,B,C] \cdots \mathsf{Body}[A,B,C'])(x,y,\Delta)$
yields the difference between the satisfying assignments of the
rule body with the old and new versions of the body predicates.
This is implemented by running two leapfrog triejoin algorithms
simultaneously, one for the rule body with the old predicates,
and one with the new versions.
The change oracle restricts evaluation to regions
matching sensitivity records.
As the maintenance rule (\ref{e:maintrule})
is evaluated, we collect sensitivity records from the leapfrog
triejoin of the new predicate versions, for use in
future maintenance.  \reffig{f:lftj3} illustrates the revised
trace, with a new satisfying assignment $x=5,y=102$ being found.

Sensitivity records are amenable to compression,
conservative approximation, and progressive refinement.
In approximating, one can aim
for high-fidelity information in volatile regions of
the database, and coarser information in static regions,
trading accuracy for parsimony to maximize overall performance.

This gives a taste for the incremental version of leapfrog
triejoin, used to repair individual rules within a transaction.
Issues such as maintaining aggregations, details
of the change oracle, etc. are explained in \cite{Veldhuizen:LB:2013}.

\section{Single transaction repair}

\label{s:graph}

\begin{figure}[t]
\subfigure[\label{f:bankrules}Two rules of a simple transaction.]{
\begin{boxedminipage}{\columnwidth}
\begin{align*}
(A) &
\begin{array}[t]{l}
\mathsf{\textasciicircum acct\_balance}[n_1]=a, \\
\mathsf{\textasciicircum acct\_balance}[n_2]=b \\
\leftarrow \begin{array}[t]{l}
      \mathsf{account\_by\_name}[\mathrm{``Alice"}]=n_1, \\
      \mathsf{account\_by\_name}[\mathrm{``Bob"}]=n_2, \\
      a = \mathsf{acct\_balance@start}[n_1] - 100, \\
      b = \mathsf{acct\_balance@start}[n_2] + 100.
      \end{array}
\end{array} \\
\\
(B) &
\begin{array}[t]{l}
\mathbf{false} \\
\leftarrow \begin{array}[t]{l}
      \mathsf{account\_by\_name}[\mathrm{``Alice"}]=n_1, \\
      \mathsf{acct\_balance}[n_1] < 0. \\
      {}
      \end{array}
\end{array}
\end{align*}
\end{boxedminipage}
} \\
\subfigure[\label{f:bankgraph}Execution graph for the rules, after some rewriting.]{
\begin{centering}
\includegraphics[scale=0.6]{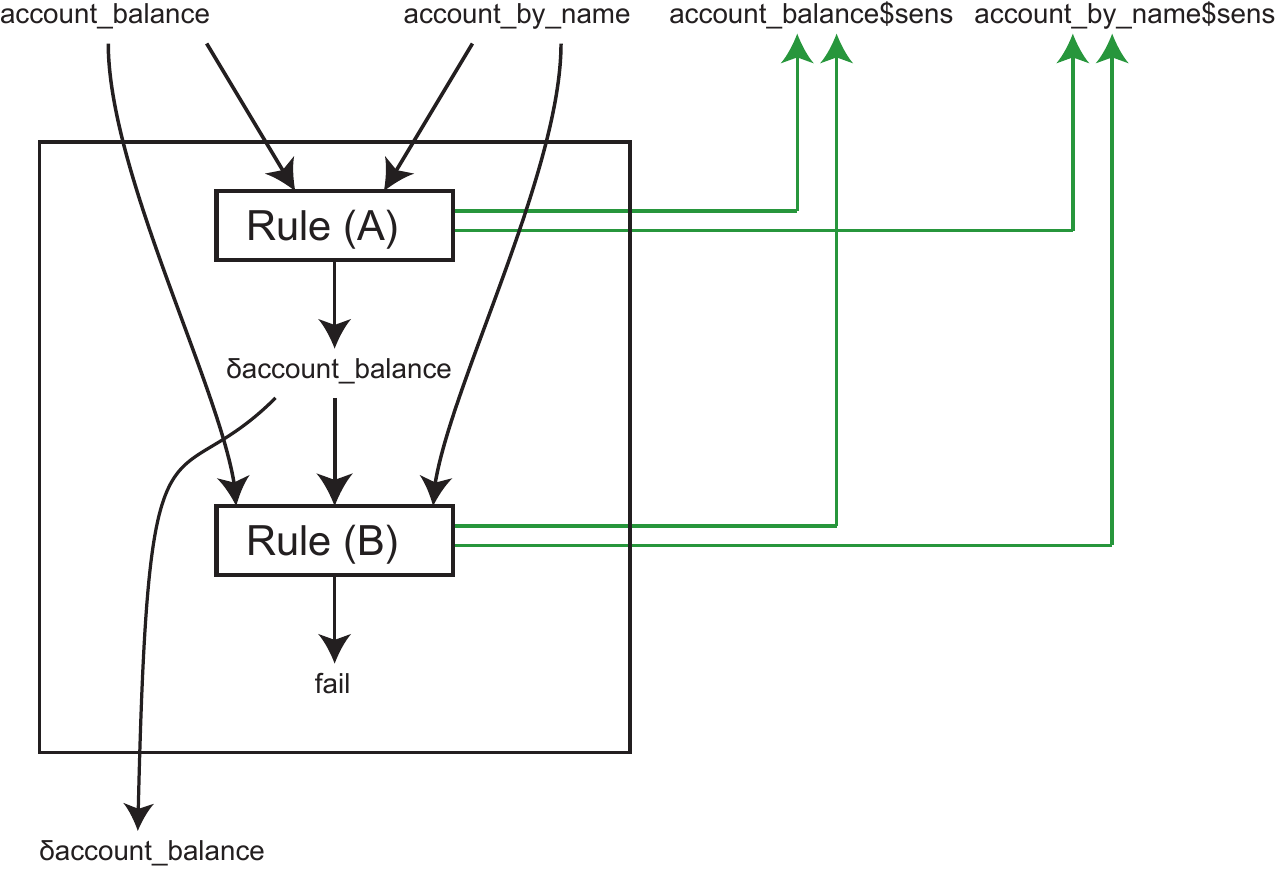}

\end{centering}

}
\caption{\label{f:bankgraphboth}A simple transaction and its execution graph.}
\end{figure}

Setting aside transactions that perform administrative functions,
schema changes, install materialized views, etc., for our purposes
a transaction is a set of rules
to be evaluated.
These rules may query database
predicates, perform computations using temporary predicates,
enforce constraints, and request changes to database predicates.
\reffig{f:bankrules} shows transaction rules 
that transfer 100 currency units
from Alice's account to Bob's account, failing if
Alice's account would be overdrawn.
In the head of rule (A) are \emph{upserts} (\textasciicircum) that
request update of
$\mathsf{acct\_balance}$ records.  The rule body retrieves account
numbers for Alice and Bob, and calculates their new account balances.
The decoration $\mathsf{@start}$ is used to distinguish the
$\mathsf{acct\_balance}$ predicate \emph{as it existed at the
start of the transaction}.  The rule (B) is a constraint:
if the balance of Alice's account \emph{at the end of the transaction}
is negative, the transaction fails.\footnote{We have omitted
complicating details.
For example, the $\mathbf{false}$
head of rule (B) would actually derive into a relation
$\mathsf{system:constraint\_fail}$, inserting a record that included
the text of the constraint rule body and the bindings of the variables.
The handling of upserts and delta predicates is more involved,
in an irrelevant way, than indicated here.
We are also omitting discussion of transaction stages,
internal rules inserted by the engine, etc.
}

The rules are placed in an \emph{execution graph}.  Rules and predicates
are vertices; there is an edge $(\mathsf{R},X)$ if predicate $X$
appears in the head of rule $\mathsf{R}$, and an edge
$(Y,\mathsf{R})$ if $Y$ appears in the body of $\mathsf{R}$.
The rules of \reffig{f:bankgraph}, after some manipulations we describe
shortly, result in the execution graph of \reffig{f:bankrules}.
The box indicates transaction scope:
database predicates appearing in rule bodies flow into the box, and deltas and
sensitivities flow out.
Rules reach the execution graph after undergoing several rounds
of rewriting.
The upserts appearing in the head of rule (A) are rewritten to
derive into a \emph{delta predicate}
$\delta \mathsf{account\_balance}[x,\Delta]=y$
containing requested changes, where
$\Delta \in \{ +, - \}$.

In single-writer operation of the database, we add a
\emph{frame rule} that applies the requested deltas.
For transaction repair, transaction rules need adjustment:
\begin{itemize}
\item Rather than having a frame rule apply the requested
deltas $\delta \mathsf{account\_balance}$ within the transaction scope,
we instead emit the deltas from the transaction without modifying
$\mathsf{account\_balance}$; these become the
\emph{delta signal} used by transaction repair.
Since the body of rule (B) refers to
the state of $\mathsf{account\_balance}$ at the \emph{end} of
the transaction, we provide rule (B) a nonmaterialized view
with the deltas applied (hence the edge to Rule (B) from
$\delta \mathsf{account\_balance}$.)
These changes make transactions purely declarative,
with no side-effects; mutation of database state only occurs 
during the apply-deltas phase of the commit process
(\refsec{s:commit}).
\item Each rule outputs sensitivity records for its body
predicates (\reffig{f:bankgraph}, green).
Sensitivity records are stripped of \emph{context keys}
(\refsec{s:inclftj}).
Sensitivities are collated at the transaction level,
forming the outgoing
\emph{sensitivity signal} used by transaction repair.
Records are never removed from the outgoing sensitivity
signal, as required for the convergence argument of
\refsec{s:convergence}.
\item Correction signals from transaction repair are handled by
providing database predicates such as 
$\mathsf{account\_balance}$ and $\mathsf{account\_by\_name}$
to the execution graph as nonmaterialized views that reflect
the database predicates after correction.  Within a rule,
we can access these corrections directly to find matching
sensitivity intervals for incremental leapfrog triejoin.
In each round of maintenance, we are not looking just at
the correction signals, but also changes made to the correction
signals.
\end{itemize}

The rules of \reffig{f:bankgraph} are implemented as
\emph{execution units}, which expose a simple interface:
$\mathsf{beginMaintenance}(\mathbf{P})$ informs an execution unit
that a round of maintenance is about to begin, and that revised versions
of the predicates named in set $\mathbf{P}$ are to be provided.
A revised predicate $P$ is provided by invoking
$\mathsf{inputChanged}(\mathsf{name},P)$.
Alternatively, one may invoke $\mathsf{inputNotChanged}(P)$.
The maintenance round ends with $\mathsf{endMaintenance}()$.
Individual rules are maintained using the incremental
leapfrog triejoin described in \refsec{s:inclftj}.

The execution graph as a whole also implements the
\emph{execution unit} interface.  When input predicates to
the graph are revised, we propagate changes through the
graph, performing rounds of maintenance on any rules
affected, and respecting dependencies.  This may result
in changes to both the outgoing delta and sensitivity
signals for the transaction.

\section{Refinements and Variations}

We now describe some variations and refinements of transaction repair.
These ideas are not yet (as of writing) implemented in our prototype, and hence
speculative.

\subsection{Selecting serialization orders.}

\label{s:arrangement}
The order in which transactions are placed
into the serialization order can have a significant impact on performance.
In general the problem falls into the broad category of
\emph{minimum linear arrangement} algorithms
\cite{Diaz:CSUR:2002}.
Some of the factors to be considered include:
\begin{itemize}
\item Read-only transactions can always be inserted at the very beginning
of the serialization order, even when there are already transactions in
flight, since their delta signals will always be null.  This avoids
any repair of read-only transactions.
\item If transaction X reads a set of data D, and transaction Y
modifies a subset of D, can be advantageous to place X before Y
in the serialization order, to avoid repair.
\item It is advantageous to group together transactions that read and write similar
sets of data in the serialization order.  This reduces the need for long
sensitivity/correction paths, which will reduce latency cost
and improve data locality when merging delta signals.
\item Define a \emph{conflict graph} $G_C=(V,E)$ whose vertices are transactions,
and there is an edge $(X,Y)$ if the sensitivities of $Y$ intersect the
deltas of $X$.  Let $G'$ be a directed, acyclic subgraph of $G_C$, constructed
by selecting a subset of transactions $V'$ and edges $E \cap V' \times V'$.
Then the transactions $V'$ can be placed in a topological order at the start
of the serialization order, without requiring any repair.
One can use this property by (a) placing new transactions in a triage
pool, where they are fully evaluated in isolation
(with an empty correction signal) to obtain their initial
sensitivities and deltas, and possibly repairing them to keep them up
to date with the latest leaf in the transaction tree; (b) building 
the conflict graph $G_C$ for the transactions in the triage pool by
determining (or conservatively approximating) conflict edges,
and (c) selecting a maximal acyclic subset of vertices $V'$ (possibly weighted by transaction
cost, or estimated repair cost) to begin a serialization order.
\item Randomization may be useful to disrupt linear chains
of conflicting transactions.  Suppose we have a sequence of
transactions $X_0,\ldots,X_{k-1}$ where $X_i$ reads a record
$r_i$ and writes a (different) record $r_{i+1}$.  This creates
a conflict chain of length $k$ whose repair could possibly be
expensive.  Suppose we choose a permutation of $X_0,\ldots,X_{k-1}$
uniformly at random for our serialization order.  The expected length
of the longest conflict chain is $2 \sqrt{k}$, a substantial reduction;
this is an instance of the `longest increasing subsequence 
of a random permutation' problem posed by Ulam.
\end{itemize}

\subsection{Load balancing techniques}

\begin{itemize}
\item
Microtransactions.  If there are large volumes of microtransactions
to be proceessed, for example, transactions that read and/or update only
a few records, it will be advantageous to group them together as if a
single transaction.  This can be accomplished by inserting a leaf node
into the transaction tree which internally contains multiple transactions
arranged in a repair circuit, but without any domain splits being
applied.  This would defer domain splits until a 
sufficiently large chunk of deltas and sensitivities have accrued.
\item Revising the domain decomposition.  In our prototype we have
so far used a static domain decomposition.  For a system that performs
well under varying workloads,
one will want to periodically revise the domain decomposition.
Ideally this would be done not according to static distribution
of data, but according to the intensity of activity, e.g., by
using random samples of recent deltas and sensitivities to
revise the domain decomposition.
\end{itemize}

\subsection{Trie surgery deltas}

Incremental leapfrog triejoin handles \emph{trie surgeries} in addition
to record-level deltas \cite{Veldhuizen:LB:2013}.
A trie surgery occurs when the first or last
record matching a key prefix is inserted or removed.
For some transactions, providing trie surgery deltas will be
necessary for performance.  To address this, we anticipate maintaining
projections of key prefixes, with support counts.  When the
$\mathsf{merge}$ operator observes a support
count making a transition from $0 \rightarrow 1$ or $1 \rightarrow 0$,
it can emit appropriate trie surgery deltas, which can be matched
against sensitivities.

\subsection{Reducing the cost of sensitivity indices.}

Collecting sensitivity information from rules in a transaction introduces
some overhead, and propagating sensitivity information through the repair
circuit can be expensive.   These costs can be reduced by employing heuristics:
\begin{itemize}
\item If we can determine statically (i.e. before initial evaluation) that
a transaction $X$ reads a set of data D, and no transaction before
$X$ modifies $D$, then we do not need to produce sensitivity records from X
for D.
\item We can trade off the precision vs. parsimony of sensitivity information.
If the sensitivity of a transaction is a subset $S \subseteq D$ where $D$
is the domain, then it is sound to use any superset $S' \supseteq S$ as the
sensitivity information.  One could, for example, report sensitivities at
the coarse level of page boundaries, or even at the level of entire predicates.
\item We can augment the circuit wiring described in \refsec{s:circuits}
with what we call \emph{sensitivity knockout} elements.
At each node of the transaction tree, insert an element that uses deltas from the left
child to "knock out" sensitivities from the right child.  This will
avoid artificial dependency chains.  For example, in a sequence of
transactions that increment a counter, the circuits would
include active correction paths from the very first transaction to the last.
Knockouts would avoid this: if transaction $2i$ wrote the counter,
and $2i+1$ read it, then the sensitivity to the counter of $2i+1$ would
be "knocked out" by the delta from transaction $2i$.  (This indicates
that changes to the counter made by transactions $< 2i$ are
irrelevant.)

Knockouts are not necessary within individual transactions
(i.e. between rules).  If a rule R1 requested an upsert
$\mathsf{\textasciicircum}F[\overline{k}]=\overline{v}$,
and rule R2 read $F[\overline{k}]$, the reported sensitivity
would be to the $\delta F$ predicate emitted by rule R1,
not to the $F$ from the database.
\end{itemize}

\subsection{Adaptations for clusters}

Transaction repair can be adapted to clusters in a straightforward way.
Suppose we have $n$ machines, each with $m$ cores, where both $n,m$
are powers of two.  (As mentioned earlier, we can pad the construction
with phantom machines and/or phantom cores processing null transactions,
to round these numbers up to the nearest power of two.)

Conceptually, we employ a transaction tree of height $\log_2(nm)-1$.  
(If needed, we can expand leaf nodes of the tree to be groups of
transactions, and not perform domain splits inside such a group.)

Each of the $n$ machines is labelled by a unique
binary string $\in \{ 0,1 \}^{(\log_2 n)-1}$.

\begin{itemize}
\item A signal $\delta_t^d$ is owned by the machine whose label is a prefix
of $td$; similarly for $s_t^d$ and $c_t^d$.  The operators emitting signals
$\delta_t^d$, $s_t^d$ and $c_t^d$ similarly reside on the machine whose label
is a prefix of $td$.  (There is
room for optimizations here; when signals traverse machine boundaries
we can consider placing the operators on either of the two machines
according to expected performance.)  Note that in \reffig{f:domainsplit},
the merge operators for sensitivities and deltas form an efficient
parallel sorting network; for example, all deltas and sensitivities
related to the leftmost subdomain are forwarded to the machine with
label $000\ldots0$, and those for the rightmost subdomain are sent to
the machine with label $111\ldots1$.
\item If a signal crosses machine boundaries, e.g., it is output by an
operator on machine $X$ and input by an operator on machine $Y$,
we transmit changes made to the signal from machine $X$ to machine $Y$.
\item When the dynamic construction of the repair circuit reaches the
last machine, and the last machine's transaction tree becomes full,
we 'wrap around,' continuing construction of the transaction tree
on the first machine.  Conceptually, one can imagine transaction
operators labelled with $\mathsf{txn}_{et}$, where $t$ is a
$\log_2(nm)-1$ binary string, and $e$ is a binary string counting the
number of times the repair circuit has wrapped around the entire
cluster.
\item Each machine can accept transactions independently; we relax
the left-to-right filling in of the transaction tree, and require this
only within each machine.  Conceptually one can think of the
transaction tree as fully constructed and containing null transactions,
which are replaced with real transactions as they arrive.
\item Each machine has its own priority queue for operators needing
refresh.  When updates to signals arrive from other machines, they
are treated the same as updates to signals within a machine; any
operators reading the signals are added to the priority queue.
\item If we use the same domain decomposition for circuits as for
dividing responsibility for durable storage of database state,
then during the apply-deltas phase of the commit process,
each machine has precisely the deltas which apply to the
subset of the database for which it is responsible.
(Note however, there is a tension here between maintaining a domain
decomposition that evenly splits the database contents vs. a decomposition
that evenly splits average activity of transactions.)
\item
The following refinement is conceptually interesting but possibly
challenging to make work in practice: suppose that, as mentioned,
the same domain decomposition is used for the repair circuit
as for responsibility of database state.  Each machine provides
its transactions with an `initial database state' consisting only
of database state currently on the machine (i.e. either state that
machine is responsible for, or cached state from other machines).
When a transaction attempts to access state unavailable on the
machine, the sensitivity signal propagates out to the responsible
machine, and the needed state is transmitted back as a correction.
This requires some elaboration to be practicable.  For example,
if a machine has no contents for a predicate $P$, and a transaction
has a rule that attempts to read $P$, leapfrog triejoin will
typically report a sensitivity for the entire domain of $P$.
This would have the unfortunate effect of fetching the entire
contents of $P$, even if only a tiny portion of $P$ was required.
One way to address this would be to respond gradually to such
requests; for example, providing the first few pages of $P$ and
waiting for more sensitivity records to be received.
\end{itemize}

\bibliography{bibliography}

\begin{thebibliography}{10}

\bibitem{Acar:SODA:2004}
Umut~A. Acar, Guy~E. Blelloch, Robert Harper, Jorge~L. Vittes, and Shan
  Leung~Maverick Woo.
\newblock Dynamizing static algorithms, with applications to dynamic trees and
  history independence.
\newblock In {\em Proceedings of the fifteenth annual ACM-SIAM symposium on
  Discrete algorithms}, SODA '04, pages 531--540, Philadelphia, PA, USA, 2004.
  Society for Industrial and Applied Mathematics.

\bibitem{Bernstein:CSUR:1981}
Philip~A. Bernstein and Nathan Goodman.
\newblock Concurrency control in distributed database systems.
\newblock {\em ACM Comput. Surv.}, 13(2):185--221, June 1981.

\bibitem{Bernstein:2009}
Philip~A Bernstein and Eric Newcomer.
\newblock {\em Principles of transaction processing}.
\newblock Morgan Kaufmann, 2009.

\bibitem{Bernstein:VLDB:2011}
Philip~A. Bernstein, Colin~W Reid, Ming Wu, and Xinhao Yuan.
\newblock Optimistic concurrency control by melding trees.
\newblock {\em Proceedings of the VLDB Endowment}, 4(11), 2011.

\bibitem{Chatterjee:SUPER:1990}
Siddhartha Chatterjee, Guy~E. Blelloch, and Marco Zagha.
\newblock Scan primitives for vector computers.
\newblock In {\em In Proceedings Supercomputing '90}, pages 666--675, 1990.

\bibitem{Chen:2005}
Songting Chen.
\newblock {\em Efficient Incremental View Maintenance for Data Warehousing}.
\newblock PhD thesis, Worcester Polytechnic Institute, 2005.

\bibitem{Diaz:CSUR:2002}
Josep D{\'i}az, Jordi Petit, and Maria Serna.
\newblock A survey of graph layout problems.
\newblock {\em ACM Comput. Surv.}, 34(3):313--356, 2002.

\bibitem{Driscoll:STOC:1986}
James~R. Driscoll, Neil Sarnak, Daniel~Dominic Sleator, and Robert~Endre
  Tarjan.
\newblock Making data structures persistent.
\newblock In {\em {ACM} Symposium on Theory of Computing}, pages 109--121,
  1986.

\bibitem{Gupta:1999}
Ashish Gupta and Iderpal~Singh Mumick.
\newblock {\em Materialized views: techniques, implementations, and
  applications}.
\newblock MIT press, 1999.

\bibitem{Gupta:SIGMOD:1993}
Ashish Gupta, Inderpal~Singh Mumick, and V.~S. Subrahmanian.
\newblock Maintaining views incrementally.
\newblock In Peter Buneman and Sushil Jajodia, editors, {\em Proceedings of the
  1993 {ACM} {SIGMOD} International Conference on Management of Data, {SIGMOD}
  '93, Washington, {DC}, May 26--28, 1993}, volume 22(2) of {\em SIGMOD Record
  (ACM Special Interest Group on Management of Data)}, pages 157--166,
  pub-ACM:adr, 1993. ACM Press.

\bibitem{Kung:TODS:1981}
Hsiang-Tsung Kung and John~T Robinson.
\newblock On optimistic methods for concurrency control.
\newblock {\em ACM Transactions on Database Systems (TODS)}, 6(2):213--226,
  1981.

\bibitem{Okasaki:1998}
Chris Okasaki.
\newblock {\em Purely Functional Data Structures}.
\newblock Cambridge University Press, Cambridge, UK, 1998.

\bibitem{Veldhuizen:LB:2013}
Todd~L. Veldhuizen.
\newblock Incremental maintenance for leapfrog triejoin.
\newblock Technical Report LB1202, LogicBlox Inc., 2013.
\newblock \href{http://arxiv.org/abs/1303.5313}{arXiv:1303.5313}.

\bibitem{Veldhuizen:ICDT:2014}
Todd~L. Veldhuizen.
\newblock Leapfrog triejoin: A simple, worst-case optimal join algorithm.
\newblock In {\em Proceedings of the International Conference on Database
  Theory (ICDT)}, March 2014.
\newblock \href{http://arxiv.org/abs/1210.0481}{arXiv:1210.0481}.

\end{thebibliography}
\bibliographystyle{plain}

\end{document}